 \documentclass[prb, twocolumn, nopacs, superscriptaddress,amsmath]{revtex4}
\usepackage{dcolumn}
\usepackage{bm}
\usepackage{graphicx}
\usepackage{color}
\usepackage{subfigure}
\usepackage{hyperref}
\usepackage{latexsym}
\usepackage{amsthm}
\usepackage{amssymb}
\usepackage{braket}
\usepackage{bbm}
\usepackage{mathbbol}
\DeclareGraphicsExtensions{.jpg,.pdf, .mps, .png, .eps, .ps, .EPS,.gif}

\def\be{\begin{equation}}
\def\ee{\end{equation}}
\def\bc{\begin{center}}
\def\ec{\end{center}}
\def\bea{\begin{eqnarray}}
\def\eea{\end{eqnarray}}
\newcommand{\avg}[1]{\langle{#1}\rangle}
\newcommand{\Avg}[1]{\left\langle{#1}\right\rangle}

\begin{document}
\title{Dirac synchronization is rhythmic and explosive}
\author{Lucille Calmon}
\affiliation{School of Mathematical Sciences, Queen Mary University of London, London, E1 4NS, United Kingdom}
\author{Juan G. Restrepo}
\affiliation{Department  of  Applied  Mathematics,  University  of  Colorado  at  Boulder,  Boulder,  CO  80309,  USA}
\author{Joaqu\'{\i}n J. Torres}
\affiliation{Departamento de Electromagnetismo y Física de la Materia and Instituto Carlos I de Física Teórica y Computacional, Universidad de Granada, 18071, Granada, Spain}
\author{Ginestra Bianconi}
\email{ginestra.bianconi@gmail.com}
\affiliation{School of Mathematical Sciences, Queen Mary University of London, London, E1 4NS, United Kingdom}
\affiliation{The Alan Turing Institute, 96 Euston Road, London,  NW1 2DB, United Kingdom}
\begin{abstract}
Topological signals defined on nodes, links and higher dimensional simplices define the dynamical state of a network or of a simplicial complex.  As such, topological signals are attracting increasing attention in network theory, dynamical systems, signal processing and machine learning. 
 Topological signals defined on the nodes are typically studied in network dynamics, while topological signals defined on links are much less explored.
Here we investigate  {Dirac} synchronization,
describing locally coupled topological signals defined on the nodes and on the links of a network, and treated using the  topological Dirac operator. The
dynamics of signals defined on the nodes is affected by a phase lag depending on the dynamical
state of nearby links and vice versa. We show that  {Dirac} synchronization on a fully connected network is explosive  with a hysteresis loop characterized by a discontinuous forward transition and a continuous backward transition. The analytical investigation of the  phase diagram  provides a theoretical understanding of this topological explosive synchronization. The model also displays an exotic coherent synchronized phase, also called rhythmic phase, characterized by non-stationary order parameters which can shed light on topological mechanisms for the emergence  of brain rhythms.
\end{abstract}
\maketitle

\section{Introduction}

Synchronization \cite{kuramoto1975,strogatz2018nonlinear,strogatz2000kuramoto,arenas2008synchronization,Boccaletti,pikovsky2003synchronization,schaub2016graph} pervades  physical and biological systems \cite{strogatz2012sync,gross2021not}. It is key to characterize  physiological \cite{glass2020clocks} and  brain rhythms \cite{buzsaki2006rhythms}, to understand collective animal behavior \cite{couzin2018synchronization}, and is also observed in  non biological systems such as coupled  Josephson junctions \cite{wiesenfeld1998frequency}, lasers \cite{soriano2013complex} and ultracold atoms \cite{zhu2015synchronization, witthaut2017classical}.

While topology is recognized to play a fundamental role in theoretical physics \cite{nakahara2003geometry,yang2019topology}, statistical mechanics \cite{kosterlitz1973ordering} and condensed matter \cite{shankar2020topological,shen2012topological,fruchart2013introduction,tang2020topology}, its role in determining the properties of  synchronization phenomena \cite{millan2021geometry,millan2020explosive,ghorbanchian2020higher,hart2019topological} has only started to be unveiled.

Recently, sparked by the growing interest in higher-order networks and simplicial complexes \cite{bianconi2021higher,battiston2021physics,battiston2020networks,bick2021higher}, the study of topology \cite{giusti2016two,otter2017roadmap,ziegler2022balanced,krishnagopal2021spectral,jost2015mathematical,taylor2015topological,mulas2020coupled,patania2017topological,petri2014homological,Jost} and topological signals \cite{millan2021geometry,millan2020explosive,torres2020simplicial,ghorbanchian2020higher,bianconi2021topological,arnaudon2021connecting,Barbarossa,schaub2020random,schaub2021signal,ebli2020simplicial,bodnar2021weisfeiler,deville2020consensus} is gaining increasing attention in  network theory, dynamical systems and machine learning. In a network, topological signals are dynamical variables that can be associated to both nodes and links.
While in dynamical systems it is common to consider dynamical variables associated to the nodes of a network and affected by interactions described by the links of the network, the investigation of the dynamics of higher order topological signals is only at its infancy\cite{millan2021geometry,millan2020explosive,torres2020simplicial,ghorbanchian2020higher,bianconi2021topological,arnaudon2021connecting,deville2020consensus}.

Topological signals associated to the links of a network are of relevance for a variety of complex systems. For instance, topological signals associated to links, also called edge signals, are attracting increased interest in the context of brain research \cite{sporns1,sporns2} and can be relevant to understand the behaviour of actual neural systems. Indeed, they can be associated to synaptic oscillatory signals related to intracellular calcium dynamics \cite{Dupont2011}, involved in synaptic communication among neurons \cite{astrocyte1997}, which could influence the processing of information through the synapses and memory storage and recall. Furthermore, edge signals can be used to model  fluxes in  biological transportation networks and even in power-grids \cite{deville_dimitri,rocks2021hidden,katifori2010damage,kaiser2020discontinuous}.

It has recently been shown \cite{millan2020explosive} that the topological signals associated to simplices of a given dimension in a simplicial complex can be treated with a higher-order Kuramoto model that uses boundary operators to show how the irrotational and solenoidal components of the signal synchronize. When these two components of the dynamics are coupled by a global adaptive coupling\cite{millan2020explosive}, the synchronization transition then becomes abrupt. These results  open the perspective to investigate the coupled dynamics of signals of different dimension both on networks and on simplicial complexes. Thus, in \cite{ghorbanchian2020higher} it has been shown that a global adaptive coupling can give rise to a discontinuous synchronization transition. However  the global adaptive mechanism adopted in Ref. \cite{ghorbanchian2020higher} is not topological and on a fully connected network, the synchronization dynamics proposed in \cite{ghorbanchian2020higher} does not display a stable hysteresis loop. Therefore,   {important open theoretical questions are} whether a discontinuous synchronization transition can be observed when topological signals are coupled   {locally}, and whether a topological coupling of the signals can lead to a stable hysteresis loop even for fully connected networks.
 
Here, we propose a dynamical model called  {Dirac} synchronization that uses topology and, in particular, the topological Dirac operator \cite{bianconi2021topological,lloyd2016quantum} to  couple locally the   dynamics of topological signals defined on the nodes and links of a network.  {Dirac} synchronization describes a Kuramoto-like dynamics for phases associated to the nodes and the links, where for the synchronization dynamics defined on the nodes, we introduce a time-dependent phase lag depending on the dynamics of the topological signals associated to the nearby links, and vice-versa. This adaptive and local coupling mechanism induces a non-trivial feedback mechanism between the two types of topological signals leading to a rich phenomenology. The main phenomena observed include a discontinuous backward transition and a   rhythmic phase where a complex order parameter oscillates at constant frequency also in the frame in which the intrinsic frequencies are zero in average.

\section{Results and Discussion}
\subsection{Motivation and Main Results}
Dirac synchronization adopts a coupling mechanism of node and link topological signals  dictated by topology that makes use of the  topological Dirac operator \cite{bianconi2021topological,lloyd2016quantum} and the higher-order Laplacians \cite{Jost,torres2020simplicial,Barbarossa}. A crucial element of Dirac synchronization is the introduction of adaptive phase lags both for phases associated to nodes and phases associated to links. Constant phase lags have been traditionally studied in the framework of the Sakaguchi and Kuramoto model \cite{sakaguchi1986soluble,english2015experimental}, which  in the presence of a careful fine tuning of the internal frequencies\cite{omel2012nonuniversal,omel2013bifurcations},  {time delays\cite{yeung1999time} or non trivial network structure \cite{eldering2021chimera}}  can lead to non-trivial phase transitions and  {chimera states}.  {Recently,} space-dependent phase lags have been considered as pivotal elements to describe cortical oscillations \cite{breakspear2010generative}. Here, we show that time-dependent phase-lags are a natural way to couple dynamical topological signals of nodes and links, leading to a very rich phenomenology and a non-trivial phase diagram,  including discontinuous synchronization transitions even for a Gaussian distribution of the internal frequencies.

The  properties of  {Dirac} synchronization are very rich and differ significantly from the properties of the higher-order Kuramoto \cite{ghorbanchian2020higher}.
 {Most importantly the order parameters of the two models are not the same, as Dirac synchronization has order parameters which are linear combinations of the signal of nodes and links revealing a very interdependent dynamics of the two topological signals. On the contrary,  the order parameters of the higher-order Kuramoto model are associated exclusively to one type of topological signals: there is one order parameter depending on  node signals and one depending on link signals.} 
Moreover, the phase diagram of Dirac synchronization on a fully connected network  includes a discontinuous forward transition and a continuous backward transition with  a thermodynamically stable hysteresis loop.  {Finally}, the coherent phase of  {Dirac synchronization} also called rhythmic phase, is non-stationary. In this phase the nodes are not just distinguished in two classes (frozen and drifting) like in the Kuramoto model but they might display a non-stationary dynamics in which one of the two phases associated to the node is drifting and the other is oscillating with a relatively small amplitude while still contributing to the corresponding order parameter.

In this work, we investigate the phase diagram of  {Dirac} synchronization numerically and analytically capturing both the stationary and the non-stationary  phases of  {Dirac} synchronization. The theoretical results are in excellent agreement with  extensive numerical simulations.

Interestingly, we can predict analytically the critical coupling constant for the discontinuous forward transition and capture the salient features of the observed rhythmic phase.

Large attention has been recently devoted to investigate which mechanisms are able to induce discontinuous, explosive synchronization transitions \cite{boccaletti2016explosivereview,d2019explosive} in simple networks  { \cite{gomez2011explosive,coutinho2013kuramoto,Boccaletti_explosive,zhang2013explosive,arola2022self}, adaptive networks \cite{avalos2018emergent}, multiplex networks \cite{Boccaletti_explosive,nicosia2017collective,peron2020collective,sarika,jalan2019inhibition,eldering2021chimera}} and simplicial complexes \cite{millan2020explosive,skardal2019abrupt,skardal2020higher,lucas2020multiorder}. 
The discontinuous transition  of  {Dirac} synchronization  is driven by the onset of instability of the incoherent phase, 
 {similarly to what happens in other synchronization models treating exclusively nodes signals \cite{peron2020collective,arola2022self}}. Moreover,  {Dirac} synchronization is driven by a topologically induced mechanism resulting in an abrupt, discontinuous synchronization transition that cannot be reduced to the recently proposed framework \cite{kuehn2021universal} intended to unify the different approaches to explosive  {transitions}.

The emergent rhythmic phase of Dirac synchronization extends to a wide range of  values of the coupling constant and is actually the only coherent phase that can be observed in the infinite network limit. The rhythmic phase, characterized by non-stationary order parameters, might shed light on the mechanisms involved in the appearance of brain rhythms and cortical oscillations \cite{buzsaki2006rhythms,breakspear2010generative,cabral2011role} since Kuramoto-like dynamics has been reported to be a very suitable theoretical framework to investigate such brain oscillatory behaviour \cite{breakspear2010generative}. Oscillations of the order parameters occur, for example, in the presence of stochastic noise and time delays  \cite{strogatz1999timedelay}, in networks with neighbour frequency correlations \cite{restrepo2014mean}, and in the context of the $D$-dimensional Kuramoto model \cite{Ott_Girvan} in\cite{dai2020discontinuous}.  However, while in~\cite{dai2020discontinuous} the magnitude of the order parameter displays large fluctuations, in Dirac synchronization we have a wide region of the phase diagram in which  one of the two complex order parameters (the complex order parameter $X_{\alpha}$) oscillates at a low frequency  without large fluctuations in its absolute value, while the other (the complex order parameter $X_{\beta}$) has non-trivial phase and amplitude dynamics. In addition to the explosive forward synchronization transition and the complex rhythmic phase described above, the bifurcation diagram of the system features a continuous backward transition, resulting in a very rich phenomenology that could potentially give insight into new mechanisms for the generation of brain rhythms \cite{breakspear2010generative}.

 {Note that Dirac synchronization is fully grounded in discrete topology but it is rather distinct from the  synchronization model proposed in  Ref.\cite{sone2022topological},  the most notable differences being (i) that  Dirac synchronization treats topological signals defined on nodes and links of a network while Ref. \cite{sone2022topological}  only treats node signals, (ii) that Dirac synchronization makes use of the topological Dirac operator which Ref. \cite{sone2022topological} does not, and (iii) that Dirac synchronization treats non-identical oscillators while Ref. \cite{sone2022topological} focuses on identical ones.}

\begin{figure*}[tbh]
\centering
\includegraphics[width=2.0\columnwidth]{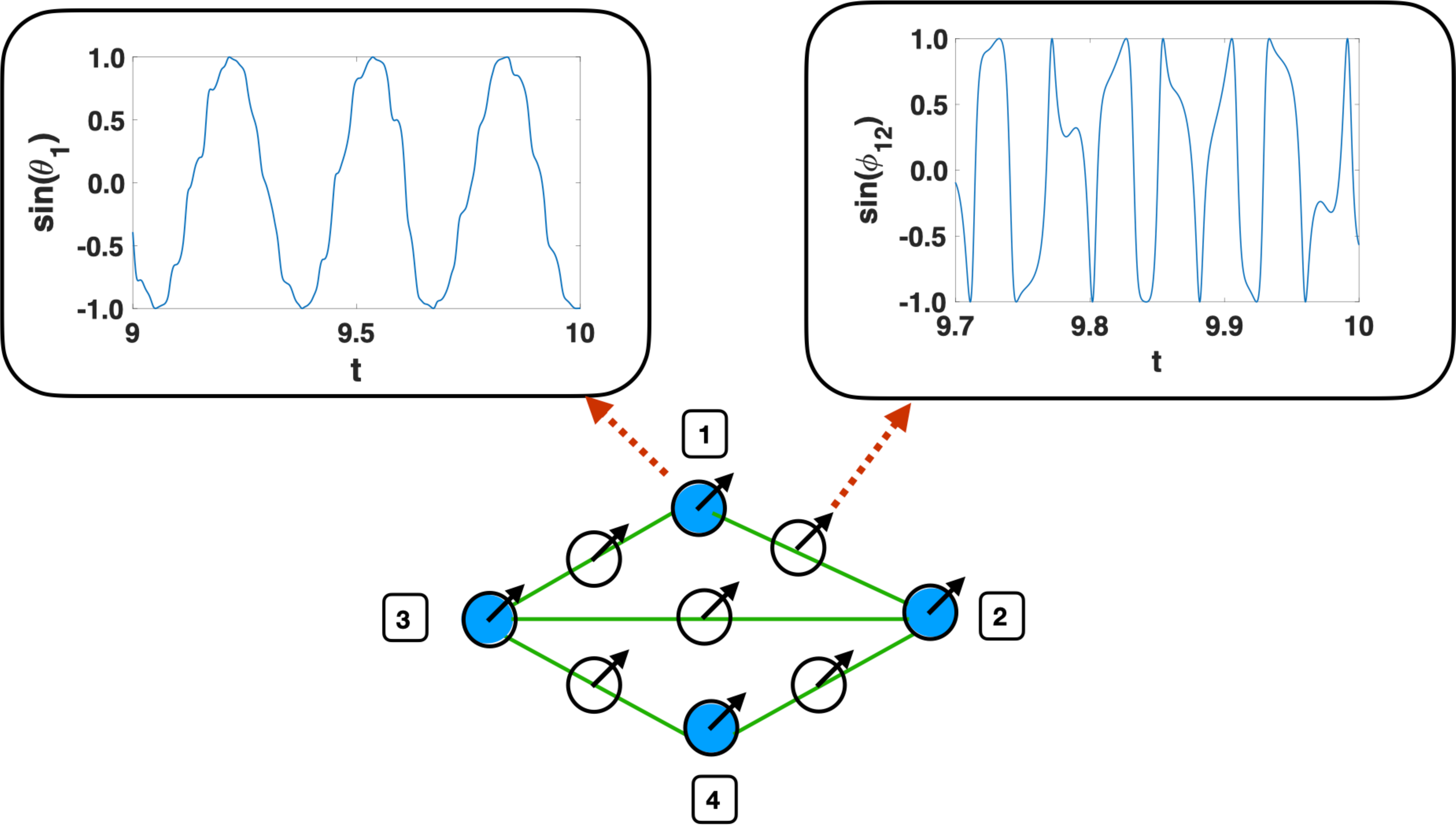}
\caption{{\bf Schematic representation of topological signals defined on the nodes and the links of a network.}
In Dirac synchronization topological signals associated to the  nodes and the links of a network are  coupled locally thanks to the Dirac operator. The considered topological signals are the phases of oscillators associated to nodes (blue oscillator symbols placed on nodes) or to links (non-filled oscillator symbols placed on links) of a network. The two insets show schematically the time-series of  node $[1]$ and  link $[1,2]$ signals respectively. }
\label{fig:1}       
\end{figure*}
\subsection{ Dirac Synchronization with local coupling-}

\subsubsection{ Uncoupled synchronization of topological signals -}
We consider a network $G=(V,E)$ formed by a set of $N$ nodes $V$ and a set of $L$ links $E$. The topology of the network is captured by the incidence matrix  ${\bf B}$ mapping any link $\ell$ of the network to its two end-nodes. Specifically, the incidence matrix ${\bf B}$   is a rectangular matrix of size $N\times L$ with elements
 {
    \begin{equation}
    [B_{1}]_{i\ell}=\left\{\begin{array}{ccc}1 & \mbox{if}\   \ell=[j,i]\  \mbox{and}\   j<i,\\ -1 & \mbox{if} \  \ell=[i,j] \  \mbox{and}\   i<j,\\  0 & \  \mbox{otherwise}\end{array}\right.
    \label{B_link}
    \end{equation}}
The standard Kuramoto dynamics \cite{kuramoto1975} describes the synchronization of the phases $\bm\theta=(\theta_1,\theta_2\ldots, \theta_N)^{\top}$ associated to the nodes of the network. In absence of interactions, each phase $\theta_i$ oscillates at some intrinsic frequency $\omega_i$, typically drawn from a unimodal random distribution. Here, we consider the normal distribution $\omega_i\sim \mathcal{N}(\Omega_0,1/\tau_0)$. However, the phases of next nearest neighbours are coupled to each other by an interaction term that tends to align phases. This term is modulated by a coupling constant $\hat\sigma$, which is the control parameter of the dynamics. In terms of the incidence matrix ${\bf B}$, the standard Kuramoto model can be expressed as  
\bea
\dot {\bm {\theta}}&=&\bm\omega -\hat\sigma {\bf B}\sin \left({\bf B}^{\top}\bm\theta\right),
\label{2}
\eea
where $\bm\omega=(\omega_1,\omega_2,\ldots, \omega_N)^{\top}$ indicates the vector of intrinsic frequencies. Note that in Eq. (\ref{2}) and in the following by $\sin ({\bf x})$, we indicate the vector where the sine function is taken 
elementwise.
As a function of the coupling constant, the Kuramoto model is known to display a synchronization transition with order parameter 
\bea
R_{\theta}=\left|\frac{1}{N}\sum_{i=1}^N e^{\textrm{i}\theta_i}\right|.
\eea
The higher-order Kuramoto model \cite{millan2020explosive} captures synchronization of topological signals (phases) defined on the $n$-dimensional faces of a simplicial complex, with $n>0$.  Let us consider the topological signals defined on the links, denoted by the vector of phases $\bm\phi=(\phi_{\ell_1},\phi_{\ell_2},\ldots, \phi_{\ell_L})^{\top}$. On a network, formed exclusively by nodes and links, the higher-order Kuramoto dynamics for these phases can be written as 
\bea
\dot{\bm\phi}=\tilde{\bm\omega}-\hat{\sigma}{\bf B}^{\top}\sin({\bf B}\bm\phi),
\eea
where  $\tilde{\bm\omega}$ indicates the vector of internal frequencies of the links, $\tilde{\bm\omega}=(\tilde{\omega}_{\ell_1},\tilde{\omega}_{\ell_2},\ldots, \tilde{\omega}_{\ell_L})^{\top}$, with $\tilde{\omega}_{\ell}\sim \mathcal{N}(\Omega_1,1/\tau_1)$.
The phases associated to the links can be projected to the nodes by applying the incidence matrix that acts like a discrete divergence of the signal defined on the links. The projection of the phases of the links onto the nodes, indicated by $\bm\psi$, is given by 
\bea
\bm\psi={\bf B}\bm \phi.
\label{projection}
\eea
As a function of the coupling constant, the higher-order Kuramoto model has been recently shown in Ref. \cite{millan2020explosive} to display a synchronization transition with order parameter 
\bea
R_{\psi}=\left|\frac{1}{N}\sum_{i=1}^N e^{\textrm{i}\psi_i}\right|.
\eea
Let us define the Dirac operator \cite{bianconi2021topological} of the network  as the $(N+L)\times(N+L)$ matrix with block structure
\bea
\boldsymbol{\mathcal{D}}=\left(\begin{array}{cc}{\bf 0} & {\bf B}\\
{\bf B}^{\top}& {\bf 0}\end{array}\right),
\eea
whose square is given by the Laplacian operator 
\bea
\boldsymbol{\mathcal{L}}=\boldsymbol{\mathcal{D}}^2=\left(\begin{array}{cc}{\bf L}_{[0]} &  {\bf 0}\\ {\bf 0}&{\bf L}_{[1]}\end{array}\right).
\label{D2}
\eea
Here ${\bf L}_{[0]}={\bf B}{\bf B}^{\top}$ is the graph Laplacian describing node to node diffusion occurring through links, and ${\bf L}_{[1]}={\bf B}^{\top}{\bf B}$ is the 1-(down)-Laplacian describing the diffusion from link to link through nodes \cite{torres2020simplicial,Barbarossa}.

Using the Dirac operator, the uncoupled dynamics of nodes and links of a network can simply be written as
\bea
\dot{\bm {\Phi}}&=&\bm\Omega -\hat\sigma \boldsymbol{\mathcal{D}}\sin \left(\boldsymbol{\mathcal{D}}\bm\Phi\right),
\label{uncoupled}
\eea
where $\bm\Phi$ and $\bm\Omega$ are $N+L$ dimensional column vectors given by 
\bea
{\bm\Phi}=\left(\begin{array}{c}\bm\theta \\
\bm\phi\end{array}\right),\ \  {\bm{\Omega}}=\left(\begin{array}{c}\bm\omega\\
\bm{\tilde{\omega}}\end{array}\right).
\eea
The dynamics of the phases associated to the nodes is identical to the standard Kuramoto dynamics, and for the vast majority of network topologies, it displays a continuous phase transition at a non-zero value of the coupling constant \cite{arenas2008synchronization}. However, the dynamics of the phases associated to the links, for the higher-order Kuramoto model, displays a continuous phase transition at zero coupling constant \cite{millan2020explosive}. 

\subsubsection{ {Dirac} synchronization}
Having defined the uncoupled dynamics of topological signals associated to nodes and links, given by Eqs.~(\ref{uncoupled}), an important  {theoretical } question that arises is how these equations can be modified to couple  {topological signals defined on nodes and links} in non-trivial ways.
In Ref.~\cite{ghorbanchian2020higher}, a global adaptive coupling modulating the coupling constant with the order parameters $R_{\theta}$ and $R_{\psi}$ was shown to lead to a discontinuous explosive transition of the coupled topological signals. However, the adaptive coupling proposed in \cite{ghorbanchian2020higher} is not  {local}: it does not admit a generalization that locally couples the different topological signals, a desirable feature since it might be argued that physical systems are typically driven by local dynamics. For instance, if the dynamics of nodes and links is assumed to treat brain dynamics, it would be easier to justify a local coupling mechanism rather than a global adaptive dynamics.
Here we formulate the equations for the  {Dirac} synchronization of  {locally} coupled signals defined on nodes and links.
We start from the uncoupled Eq. (\ref{uncoupled}) and introduce a local adaptive term in the form of a phase lag. More precisely, we introduce a phase lag for the node dynamics that depends on the topological signal associated to the nearby links, and vice versa, we consider a phase lag for the link dynamics that depends on the signal on the nodes at its two endpoints.
 The natural way to introduce these phase lags is by using the Laplacian matrix $\boldsymbol{\mathcal{L}}$, with an appropriate normalization to take into consideration the fact that nodes might have very heterogeneous degrees, while links are always only connected to the two nodes at their endpoints.
While the model can be easily applied to any network,  {we consider the case of a fully connected network to develop a thorough theoretical treatment of the dynamics}.  As we will show, even in this case, the model displays a rich phenomenology.
Therefore, we propose the Dirac synchronization model driven by the dynamical equations  
\bea
\dot{\bm\Phi}={\bm \Omega}-\frac{\sigma}{N} \boldsymbol{\mathcal{D}}\sin (\boldsymbol{\mathcal{D}}{\bm\Phi}-{\bm\gamma}\boldsymbol{\mathcal{K}}^{-1}\boldsymbol{\mathcal{L}}{\bm\Phi}),
\label{local}
\eea
where the matrix $\boldsymbol{\mathcal{K}}$ is given by 
\bea
 \boldsymbol{\mathcal{K}}=\left(\begin{array}{cc}{\bf K}_{[0]} &  {\bf 0}\\ {\bf 0}&{\bf K}_{[1]}\end{array}\right),
\eea
i.e., $\boldsymbol{\mathcal{K}}$ is  a block-diagonal matrix whose non-zero blocks are formed by the diagonal  matrix of node degrees ${\bf K}_{[0]}$ and by the diagonal matrix ${\bf K}_{[1]}$ of link generalized degrees, encoding the number of nodes connected to each link. Therefore ${\bf K}_{[1]}$ has all diagonal elements given by $2$ in any network while ${\bf K}_{[0]}$ has all diagonal elements given by $N-1$ in the case of a fully connected network. 
Moreover, in Eq. (\ref{local}) and in the following we will make use of the matrices $\bm\gamma$ and $\boldsymbol{\mathcal{I}}$  given by 
\bea
\bm\gamma=\left(\begin{array}{cc}{\bf I}_N &  {\bf 0}\\ {\bf 0}&-{\bf I}_L\end{array}\right)\ \boldsymbol{\mathcal{I}}=\left(\begin{array}{cc}{\bf I}_N &  {\bf 0}\\ {\bf 0}&{\bf I}_L\end{array}\right),
\eea
where ${\bf I}_X$ indicates the identity matrix of dimension $X\times X$.
 {On a sparse network, Dirac synchronization obeying Eq.(\ref{local}) involves a local coupling  of the phases on the nodes with a topological signal defined on nearby links and a coupling of the phases of the links with a topological signal defined on nearby nodes. In particular we substitute the argument of the $\sin(x)$ function in Eq.(\ref{uncoupled}) with 
\bea
\boldsymbol{\mathcal{D}}\Phi\to \boldsymbol{\mathcal{D}}{\bm\Phi}-{\bm\gamma}\boldsymbol{\mathcal{K}}^{-1}\boldsymbol{\mathcal{L}}{\bm\Phi}.
\label{sub}
\eea
Indeed, to have a meaningful model, one must require that the interaction term (in the linearized system) is positive definite which for us implies that the first order term to couple the signal of nodes and links includes a phase-lag proportional to the Laplacian, i.e, proportional to the square of the Dirac operator $\boldsymbol{\mathcal{D}^2}=\boldsymbol{\mathcal{L}}$. However one could also envision more general models where higher powers of  the Dirac operator could be included.
It is to be mentioned that the introduction of the matrix $\bm\gamma$ is necessary to get a non-trivial phase diagram, while $\mathcal{K}^{-1}$ is important  to have a linearized coupling term that is semidefinite positive.
Finally, we note that the introduction of quadratic terms in Eq. (\ref{sub})  is in line with the analogous generalization of the Dirac equation for  topological insulators proposed in Ref.~\cite{shen2012topological} which also includes an additional term proportional to the square of the momentum (analogous to our Laplacian).}

 {For our analysis on fully connected networks,} we draw the intrinsic frequencies of the nodes and of the links, respectively, from the distributions $\omega_i\sim \mathcal{N}(\Omega_0,1)$ and $ \tilde{\omega}_{\ell}\sim \mathcal{N}(0,1/\sqrt{N-1})$.  {This rescaling of the frequencies on the links ensures that the induced frequencies on the nodes defined shortly in Eq. (\ref{hatomega}) are themselves normally distributed, with zero mean and unit-variance.}
It is instructive to write Eq.~(\ref{local}) separately as a dynamical system of equations for the phases $\bm\theta$ associated to the nodes and the phases $\bm\phi$ associated to the links of the network, giving 
\bea
\dot {\bm {\theta}}&=&\bm\omega -\frac{\sigma}{N} {\bf B}\sin \left({\bf B}^{\top}\bm\theta+{\bf K}_{[1]}^{-1}{\bf L}_{[1]}\bm{\phi}\right),\nonumber \\
\dot {\bm{\phi}}&=&\tilde{\bm{\omega}}-\frac{\sigma}{N}  {\bf B}^{\top}\sin \left(
{\bf B}\bm{\phi}-{\bf K}^{-1}_{[0]}{\bf L_{[0]}}\bm{\theta}
\right).
\eea
This expression reveals explicitly that the coupling between topological signals defined on nodes and links consists of adaptive phase lags determined by the local diffusion properties of the coupled dynamical signals.

Using Eq.~(\ref{D2}), we observe that the linearized version of the proposed  {dynamics} in Eq.~(\ref{local}) still couples nodes and links according to  the dynamics
 {\bea
\dot{\bm\Phi}={\bm \Omega}-\frac{\sigma}{N} (\boldsymbol{\mathcal{I}}+\bm\gamma\boldsymbol{\mathcal{D}}\boldsymbol{\mathcal{K}}^{-1})\boldsymbol{\mathcal{L}}{\bm\Phi}.
\eea
Note that one can interpret this linearized dynamics as a coupling between node phases and phases of nearby nodes (which is described by the Laplacian operator $\boldsymbol{\mathcal{L}}$). Additionally, the phases of the nodes are coupled with the phases of their incident links and of the links connected to their neighbour nodes 
(which is mediated by the term proportional to $\boldsymbol{\mathcal{D}}\boldsymbol{\mathcal{K}}^{-1}\boldsymbol{\mathcal{L}}$). A similar interpretation is in place for the dynamics of the links.
}

\subsubsection{ Dynamics projected on the nodes-}
Let us now investigate the dynamical equations that describe the coupled dynamics of the phases  {$\bm\theta$} associated to the nodes and the projection $\bm\psi$ of the phases associated to the links, with $\bm\psi$ given by Eq.~(\ref{projection}). Since $K_{[1]}$ and ${\bf B}^{\top}$  commute,  in terms of the phases $\bm\theta$ and $\bm\psi$, the equations dictating the dynamics of Dirac synchronization read
\bea
\dot {\bm {\theta}}&=&\bm\omega -\frac{\sigma}{N} {\bf B}\sin \left({\bf B}^{\top}(\bm {\theta}+\bm{\psi}/2)\right),\nonumber \\
\dot {\bm{\psi}}&=&\hat{\bm{\omega}}-\frac{\sigma}{N}  {\bf L}_{[0]}\sin \left(\bm{\psi}-{\bf K}^{-1}_{[0]}{\bf L_{[0]}}\bm{\theta}
\right),\label{topomodel}
\eea
where 
\bea
\hat{\bm{\omega}}={\bf B}\tilde{\bm{\omega}},
\label{hatomega}
\eea (see Appendix \ref{Ap0} for details on the distribution of $\hat{\bm\omega}$), and where we have used the definition of ${\bf L}_{[1]}={\bf B}^{\top}{\bf B}$. 
The  Eqs. (\ref{topomodel}) can be written {elementwise} as 
\bea
\dot {\theta}_i&=&\omega_i +\frac{\sigma}{N} \sum_{j=1}^N \sin \left(\alpha_j-\alpha_i\right),\nonumber \\
\dot {\psi}_{i}&=&\hat{\omega}_{i}+\sigma \sin (\beta_i)-\frac{\sigma}{N} \sum_{j=1}^N \sin \left(\beta_j\right),
\label{dyn_fc}
\eea
where the variables $\alpha_i$ and $\beta_i$ are defined as 
\bea
\alpha_i&=&\theta_i+\psi_i/2,\nonumber \\
\beta_i&=& \bar{c}(\theta_i-\hat{\Theta})-\psi_i,
\eea
with $\hat{\Theta}$ given by
\bea
\hat{\Theta}=\frac{1}{N}\sum_{i=1}^N \theta_{i}
\eea
and $\bar{c}=N/(N-1)$.
We observe that $\hat{\Theta}$ is the average phase of the nodes of the network that  evolves in time at a constant frequency $\hat{\Omega}$, determined only by the intrinsic frequencies of the nodes. In fact, using Eqs. (\ref{dyn_fc}), we can easily show that
\bea
\frac{d\hat{\Theta}}{dt}=\frac{1}{N}\sum_{i=1}^N{\dot{\theta}_i}=\frac{1}{N}\sum_{i=1}^N {\omega_i}=\hat{\Omega}.
\eea
Here and in the following, we indicate with $G_0(\omega)$ the distribution of  {the} intrinsic frequency $\omega$ of each node, and with $G_1(\hat{\omega})$ the marginal distribution of the frequency $\hat{\omega}$ for any generic node of the fully connected network (for  the explicit expression of $G_1(\hat{\omega})$, see Appendix \ref{Ap0}).

In order to study the dynamical Eqs.~(\ref{dyn_fc}), entirely capturing the topological synchronization on a fully connected network,
we introduce the complex order parameters $X_{\alpha}$ and $X_{\beta}$ associated to the phases $\alpha_i$ and $\beta_i$ of the nodes of the network, i.e.
 {\bea
X_{\alpha}=R_{\alpha}e^{\textrm{i}\eta_{\alpha}}&=&  \frac{1}{N}\sum_{j=1}^N e^{\textrm{i}\alpha_j},\\
X_{\beta}=R_{\beta}e^{\textrm{i}\eta_{\beta}}&=& \frac{1}{N} \sum_{j=1}^N e^{\textrm{i}\beta_j},
\eea}
where  {$R_{\xi}$ and  {$\eta_{\xi}$} are real, and $\xi\in \{\alpha,\beta\}$}.
Using this notation, Eqs. (\ref{dyn_fc}) can also be written as 
\bea
\dot {\theta}_i&=&\omega_i +\sigma \mbox{Im}\left[e^{-\textrm{i}\alpha_i}X_{\alpha}\right],\nonumber \\
\dot {\psi}_{i}&=&\hat{\omega}_{i}-\sigma \mbox{Im} X_{\beta}-\sigma  \mbox{Im} e^{-\textrm{i}\beta_i}.\label{thetaphi3}
\eea
Since Eqs.  $(\ref{thetaphi3})$ 
are invariant under translation of the $\alpha_i$ variables,
we consider the transformation 
 {\bea
\alpha_i\to \alpha_i-\hat{\Omega}t-\alpha_0
\eea
where $\alpha_0$ is independent of time. Interestingly, this invariance guarantees that if $X_{\alpha}$ is stationary then we can always choose $\alpha_0$ such that $X_{\alpha}$ is also real, i.e.,  $X_{\alpha}=R_{\alpha}$.
Independently of the existence or not of a stationary solution, this invariance can be used to also simplify} Eqs. (\ref{thetaphi3})  to 
\bea \label{angledot}
\dot{\bm{\alpha}}_i={\bm \kappa}_i+\sigma \mbox{Im}\left[{\bf \hat{X}}e^{-\textrm{i}{\bm \alpha}_i}\right],
\eea
where we indicate with  $\bm\alpha_i$ and $e^{\textrm{-i}\bm\alpha_i}$ the vectors
\bea
{\bm{\alpha}}_i=\left(\begin{array}{c} {\alpha}_i \\ {\beta}_i\end{array}\right),\quad e^{-\textrm{i}{\bm \alpha}_i}=\left(\begin{array}{c}e^{-\textrm{i}\alpha_i}\\ e^{-\textrm{i}\beta_i}\end{array}\right),
\eea
and where the vector $\bm\kappa_i=(\kappa_{i,\alpha},\kappa_{i,\beta})^{\top}$  and  the matrix ${\bf \hat{X}}$ in Eq. (\ref{angledot})  are given by  
\bea
{\bm \kappa}_i&=&\left(\begin{array}{c} \omega_i-\hat{\Omega}+\hat{\omega}_i/2-\sigma  \mbox{Im} X_{\beta}/2\\  \bar{c}\omega_i-\bar{c}\hat{\Omega}-\hat{\omega}_i+\sigma  \mbox{Im} X_{\beta}\end{array}\right),\nonumber \\
{\bf \hat{X}}&=&\left(\begin{array}{cc}X_{\alpha}&-1/2\\
\bar{c}X_{\alpha}& 1\end{array}\right).
\eea

\subsection{Phase diagram of Dirac synchronization }

Let us summarize the main properties of the phase diagram of Dirac synchronization as predicted by our theoretical derivations detailed in the Methods section.
In Dirac synchronization, the dynamics of the phases associated to the nodes is a modification of the standard Kuramoto model \cite{kuramoto1975,strogatz2000kuramoto,Boccaletti}, and includes a phase-lag that depends on the phases associated to the links.
It is therefore instructive to compare the phase diagram of Dirac synchronization with the phase diagram of the standard Kuramoto model.  
The standard Kuramoto model with normally distributed internal frequencies  {with unitary variance},  has a continuous phase transition at the coupling constant $\sigma=\sigma_c^{\mbox{std}}$ given by 
\bea
\sigma_c^{\mbox{std}}=1.59577\ldots
\label{sigmac}
\eea
with the order parameter $R_{\theta}$ becoming positive for $\sigma>\sigma_c^{\mbox{std}}$. The synchronization threshold $\sigma_c^{\mbox{std}}$ coincides with the onset of the instability of the incoherent phase where $R_{\theta}=0$. Consequently, the forward and backward synchronization transitions coincide. 
The phase diagram for Dirac synchronization is much richer.
\begin{figure}[tbh]
\centering
\includegraphics[width=1.0\columnwidth]{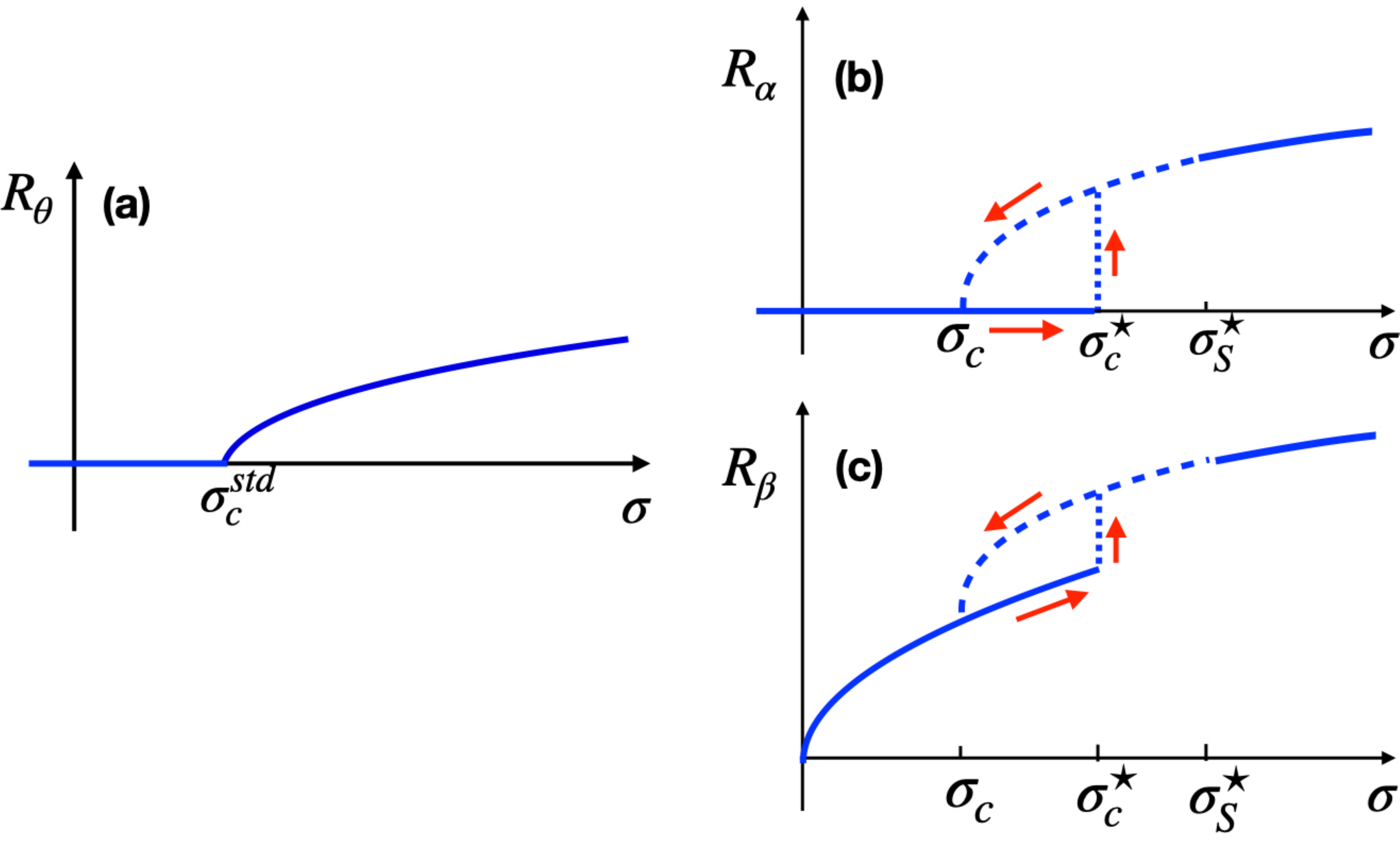}
\caption{{\bf Schematic phase diagram of Dirac synchronization.} The schematic representation of the phase diagram of the standard Kuramoto model with order parameter $R_{\theta}$ depending on the coupling constant $\sigma$ (panel (a)) is compared with the phase diagram of Dirac synchronization characterized by the two  order parameters $R_{\alpha}$ and $R_{\beta}$ depending on the coupling constant $\sigma$ (panels (b) and (c)).
The synchronization transition of the standard Kuramoto model occurs continuously at the synchronization threshold $\sigma_c^{\mbox{std}}$ and the forward and backward transitions coincide. For Dirac synchronization, the forward transition observed for increasing values of $\sigma$ starting from $\sigma=0$, is discontinuous at $\sigma_c^{\star}$ where we observe an abrupt transition from an incoherent state where $R_{\alpha}=0$ (but $R_{\beta}>0$) to a non-stationary coherent state  {of the complex order parameters} with $R_{\alpha}>0$ (and $R_{\beta}>0$).  {This non-stationary coherent phase, also called rhythmic phase, is characterized by non-trivial oscillations of the  order parameter $X_{\alpha}$ in the complex plane which are observed at essentially constant absolute value  $|X_{\alpha}|$. This phenomenon occurs in the region indicated here by dashed-lines.} For $\sigma>\sigma_S^{\star}$, where $\sigma_S^{\star}$ diverges in the limit of infinite network size, the coherent state becomes stationary. The backward transition of Dirac synchronization, observed for decreasing values of $\sigma$, first displays a transition from a stationary coherent state to a non-stationary coherent state at $\sigma_S^{\star}$, then displays a continuous phase transition from the non-stationary coherent phase to the incoherent phase at $\sigma_c$.}
\label{fig:bifurcation}       
\end{figure}
The major differences between Dirac synchronization and the standard Kuramoto model are featured in the very rich phase diagram of Dirac synchronization  {(see schematic representation in Figure $\ref{fig:bifurcation}$)} which includes discontinuous transitions, a stable hysteresis loop, and a non-stationary coherent phase (which we call rhythmic phase), in which the complex order parameters $X_{\alpha}$ and $X_{\beta}$ are not stationary.
In order to capture the phase diagram of the explosive Dirac synchronization, we need to theoretically investigate the rhythmic phase where the density distribution of the node's phases is non-stationary.

Moreover, our theoretical predictions indicate  that the onset of the instability of the incoherent state does not coincide with the bifurcation point at which the coherent phase can be first observed. This leads to a hysteresis loop characterized by a discontinuous forward transition, and a continuous backward transition.
In the context of the standard Kuramoto model, the study of the stability of the incoherent phase puzzled  the scientific community for a long time \cite{kuramoto1987statistical,kuramoto1989onset}, until Strogatz and Mirollo proved in Ref. \cite{Strogatz_Mirollo} that $\sigma_c$ corresponds to the onset of instability of the incoherent phase, and later Ott and Antonsen \cite{Ott_Antonsen} revealed the underlying one-dimensional dynamics of the order parameter in the limit $N \to \infty$. Here we conduct a stability analysis of the incoherent phase (see Appendix $\ref{crit}$ for the analytical derivations), and we find that the incoherent phase becomes unstable only for $\sigma=\sigma_c^{\star}$, with $\sigma_c^{\star}$ given by 
\bea
\sigma_c^{\star}= 2.14623\ldots.
\label{sigmacstar}
\eea
Interestingly, the bifurcation point where the synchronized branch merges with the incoherent solution in the backward transition, (see Fig.~\ref{fig:bifurcation}(b)), is expected for a smaller value of the coupling constant $\sigma_c<\sigma_c^{\star}$ that we estimate with an approximate analytical derivation, (which will be detailed below), to be equal to 
\bea
\sigma_c\simeq 1.66229\ldots.
\eea
It follows that Dirac synchronization displays a discontinuous forward transition at $\sigma_c^{\star}$, and a continuous backward transition at $\sigma_c$.

\begin{figure}[t]
\centering
\includegraphics[width=1.0\columnwidth]{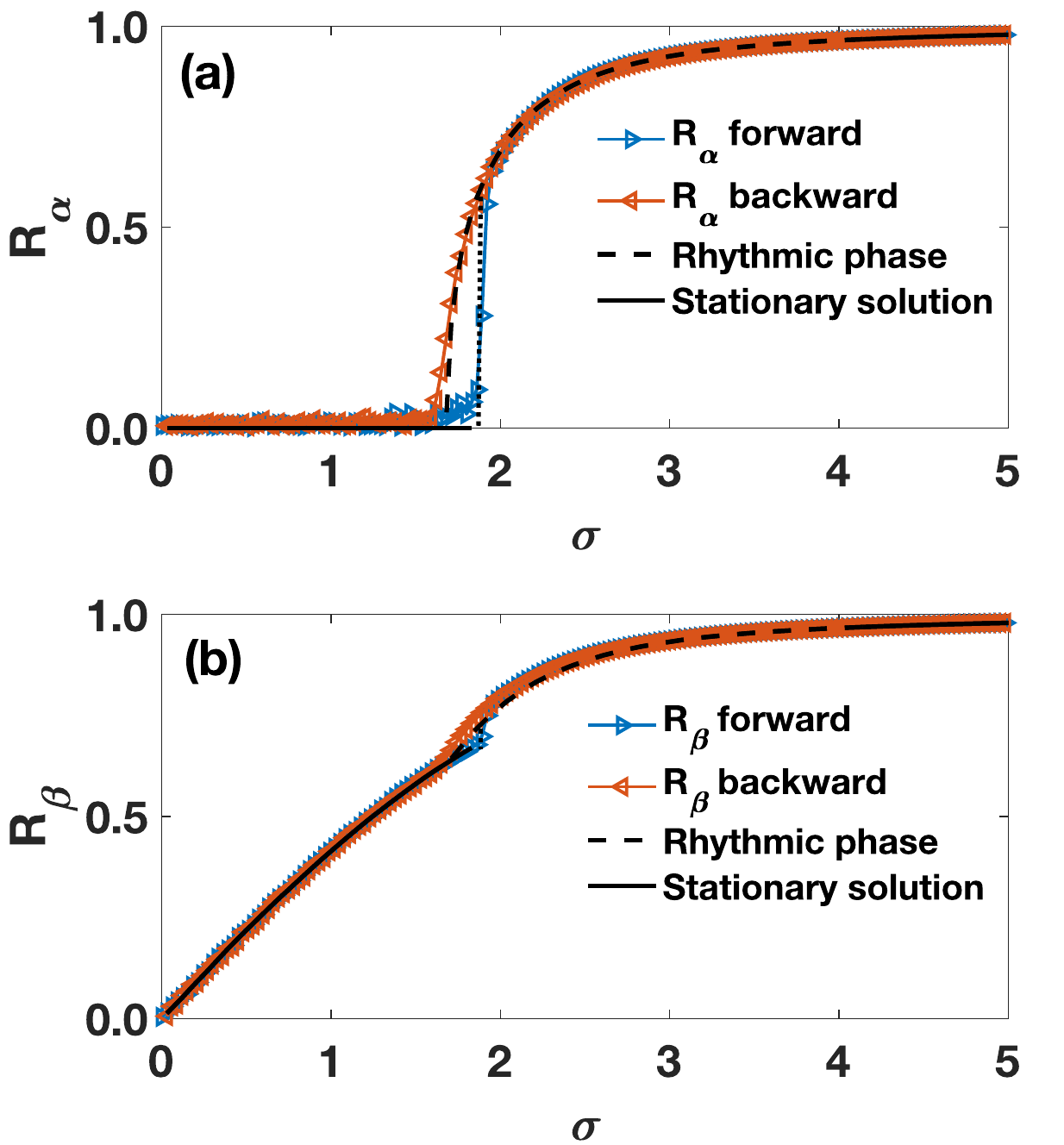}
\caption{{\bf The phase diagram of Dirac synchronization: numerical results and theoretical expectations.} The forward and backward Dirac synchronization transitions of the real order parameters $R_{\alpha}$ (panel (a)) and $R_{\beta}$ (panel (b)) are plotted as a function of the coupling constant $\sigma.$ The numerical results are obtained for a network of $N=20,000$ nodes, by integrating the dynamical equations with a 4th order Runge-Kutta method with time step $\Delta t=0.005$ where, for each value of $\sigma$ the dynamics is equilibrated up to time $T_{\text{max}}=10$. The coupling constant $\sigma$ is adiabatically increased and then decreased with steps of size $\Delta \sigma=0.03$.  For each step in $\sigma$, the plotted values of the real order parameters are averaged over the last fifth of the time series.  Black  lines indicate the theoretical predictions as developed in Eqs.  (\ref{comp2ns}). Solid lines indicate steady state solutions of the continuity equations and dashed lines represent theoretical predictions in the non-stationary coherent phase.  {The discontinuous transition point occurs at a coupling strength below the theoretical estimate due to finite size effects. We show with the dotted line the location of the  onset of the instability as extracted from finite size scaling corresponding to the network size $N=20000$.}}
\label{fig:2}     
\end{figure}

The forward transition displays a discontinuity at $\sigma_c^{\star}$, where we observe the onset of a rhythmic phase, the non-stationary coherent phase of Dirac synchronization, characterized by oscillations of the complex order parameters $X_{\alpha}$ and $X_{\beta}$. This phase persists up to a value $\sigma_S^{\star}$  where the coherent phase becomes stationary.
The backward transition is instead continuous. One observes first a transition between the stationary coherent phase and the non-stationary coherent phase at $\sigma_S^{\star}$, and subsequently a continuous transition at $\sigma_c$.

From our theoretical analysis, we predict the phase diagram sketched in Figure $\ref{fig:bifurcation}$, consisting of a thermodynamically stable hysteresis loop, with a discontinuous forward transition at $\sigma_c^{\star}$ and a continuous backward transition at $\sigma_c$. The coherent rhythmic phase disappears at $\sigma_S^{\star}$, where $\sigma_S^{\star}$ diverges in the large network limit. Therefore, for $N\to \infty$, the system always remains in the rhythmic phase.

\begin{figure*}[htbp!]
\centering
\includegraphics[width=1.7\columnwidth]{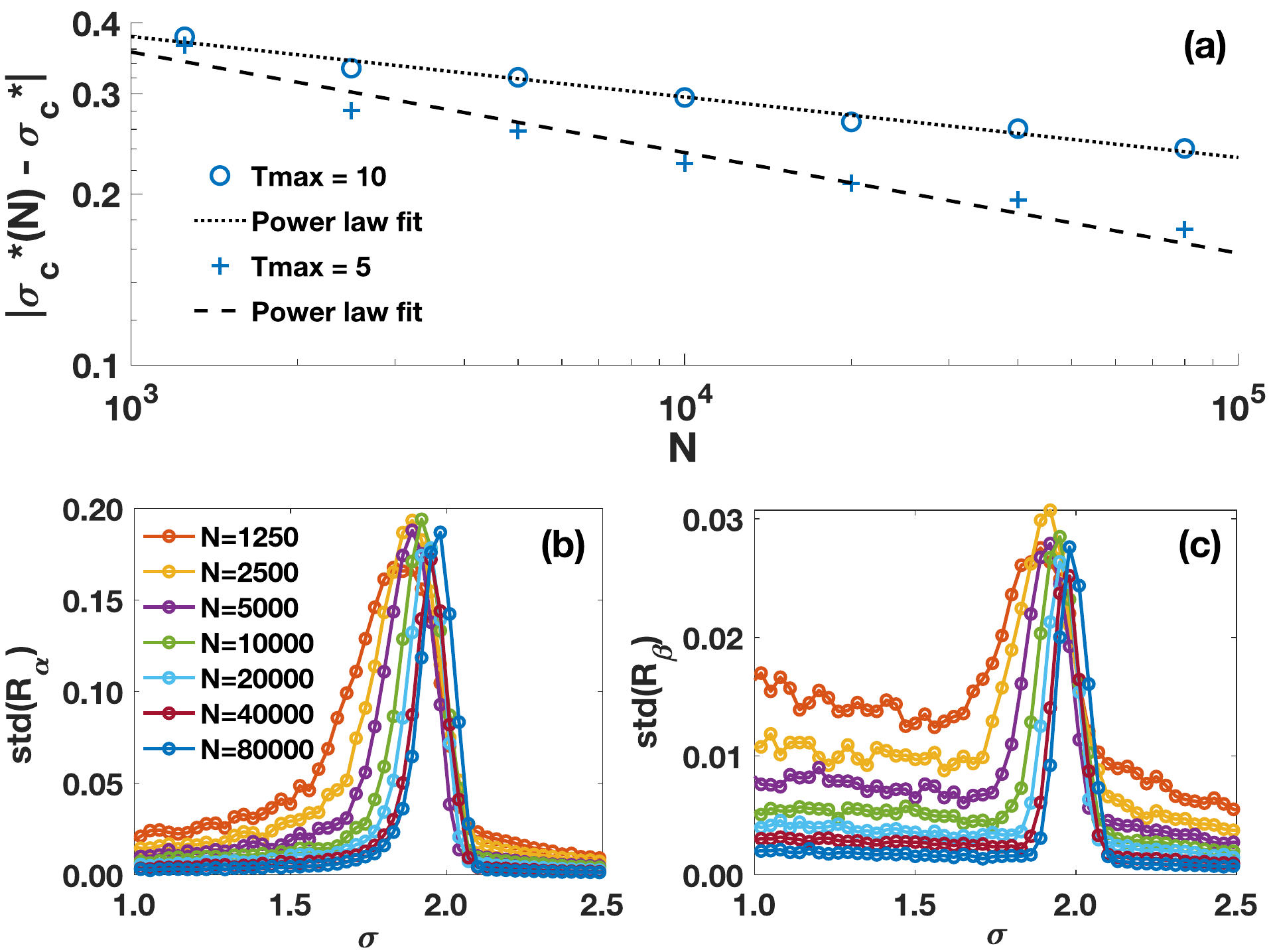}
\caption{ {{\bf The finite size effects on $\sigma_c^{\star}$.} Panel (a) shows the absolute value of the difference between the synchronization threshold $\sigma_c^{\star}(N)$ in a network of $N$ nodes and the theoretical prediction $\sigma_c^{\star}$ of the synchronization threshold in an infinite network. The value of $\sigma_c^{\star}(N)$ shown are averaged over $100$ independent realizations of the forward transition.  We find that these finite size effects depend on the equilibration time $T_{\text{max}}$ (here shown for $T_{\text{max}} = 5$ and $10$).  Indeed, for larger integration times, the probability that finite-size fluctuations about the incoherent size bring the system to the basin of attraction of the rhythmic phase increases. Thus, the observed transition point depends on $N$ (larger $N$ implies smaller fluctuations), and $T_{\text{max}}$ (larger $T_{\text{max}}$ implies larger probability of transition before reaching $\sigma_c^*$).  This process has negligible effect on the rest of the phase diagram.} The finite size effects are fitted by a power-law scaling function $|\sigma_c^{\star}(N)-\sigma_c^{\star}|=cN^{-\xi}$ with  $c = 1.208$, $\xi = 0.177$ and $c = 0.7906$, $\xi = 0.1065$ respectively for $T_{\text{max}}= 5$ (dashed line) and $T_{\text{max}}=10$ (dotted line). The standard deviation (over the $100$ iterations, $T_{\text{max}}=5$) of the order parameters $R_\alpha$ and $R_\beta$ are shown in the forward transition in panels (b) and (c) respectively. This is highest closest to the transition point, and tends to the theoretical estimate as $N$ increases.}
\label{fig:5_0}      
\end{figure*}

These theoretical predictions are confirmed by extensive numerical simulations (see Figure $\ref{fig:2}$) of the model defined on a fully connected network of  $N=20,000$ nodes (although we can observe  some minor instance-to-instance differences). These results are obtained by integrating Eqs.~(\ref{dyn_fc}) using the $4^{\text{th}}$ order Runge-Kutta method with time step $\Delta t=0.005$. 
The coupling constant $\sigma$ is first increased and then decreased adiabatically in steps of $\Delta \sigma  = 0.03$. From this numerically obtained bifurcation diagram, we see that our theoretical prediction of the phase diagram (solid and dashed lines) matches very well the numerical results (red and blue triangles). However, the discontinuous phase transition, driven by deviations from the incoherent phase, is observed at $\sigma_c^\star(N) < 2.14623...$ in finite networks. We study this effect quantitatively by measuring the transition threshold on systems of varying size $N$, averaging $100$ independent iterations for each $N$ (see Figure $\ref{fig:5_0}$). We find that the observed distance from the theoretical critical point given in Eq. (\ref{sigmacstar}) decreases consistently with a power-law in N, with scaling exponent $0.177$ (computed with integration time $T_{\text{max}}=5$), confirming further our theoretical prediction of $\sigma_c^\star$. The observed behavior is consistent with the earlier transition being caused by finite $N$ effects. For finite $N$, the system will have fluctuations about the incoherent state which may bring the system to the basin of attraction of the rhythmic phase and cause a transition even when the incoherent state is stable. Thus, the observed transition point depends on $N$ (larger $N$ implies smaller fluctuations), and $T_{\text{max}}$ (larger $T_{\text{max}}$ implies larger probability of transition before reaching $\sigma_c^*$).

Finally, we find that this rich phase diagram is only observable by considering the correct order parameters for Dirac synchronization, $R_{\alpha}$ and $R_{\beta}$, which characterize the synchronization of the coupled topological signals  {as predicted by the analytical solution of the model. This is due to the local coupling introduced in Dirac synchronization, which couples together the phases of nodes and adjacent links locally and topologically. Interestingly this  is a phenomenon that does not have an equivalent in the model \cite{ghorbanchian2020higher}  in which  the signals of nodes and links are coupled by a global order parameter}. 
Therefore  {as we will show with our theoretical derivation of the phase diagram of the model, the  phases $\bm \alpha$ and $\bm \beta$ become the relevant variables to consider instead of the original phases associated exclusively to the nodes $\bm \theta$ and to the projected signal of the links into the nodes $\bm \psi$.} Indeed,  {in agreement with our  theoretical expectations},  numerical results clearly show that the transition cannot be detected if one considers the na\"ive uncoupled order parameters $R_{\theta}$ and $R_{\psi}$, which remain close to $0$ for all values of $\sigma$.

\begin{figure*}[htbp!]
\centering
\includegraphics[width=1.7\columnwidth]{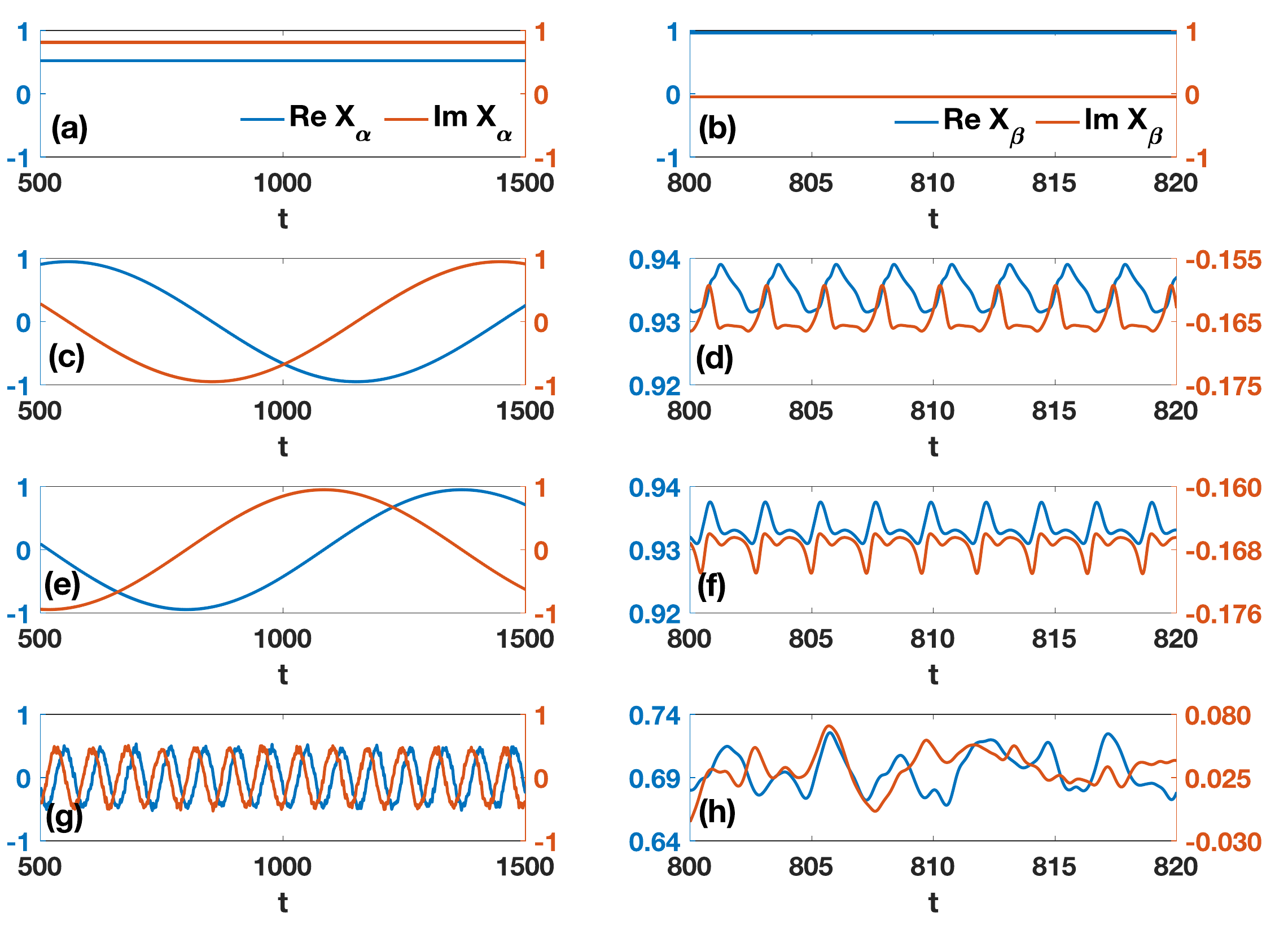}
\caption{{\bf Post-transient time evolution of the real and imaginary parts of the parameters $X_{\alpha}$ and $X_{\beta}$.} Numerical results are shown for a network of size $N=500$ during the downward transition. These results are obtained with $\sigma=4$ (panels (a) and (b)), $\sigma=3.4$ (panels (c) and (d)), for $\sigma=3.37$ (panels (e) and (f)) and for $\sigma=1.69$ (panels (g) and (h)).}
\label{fig:4}       
\end{figure*}

\begin{figure*}[htbp!]
\centering
\includegraphics[width=1.80\columnwidth]{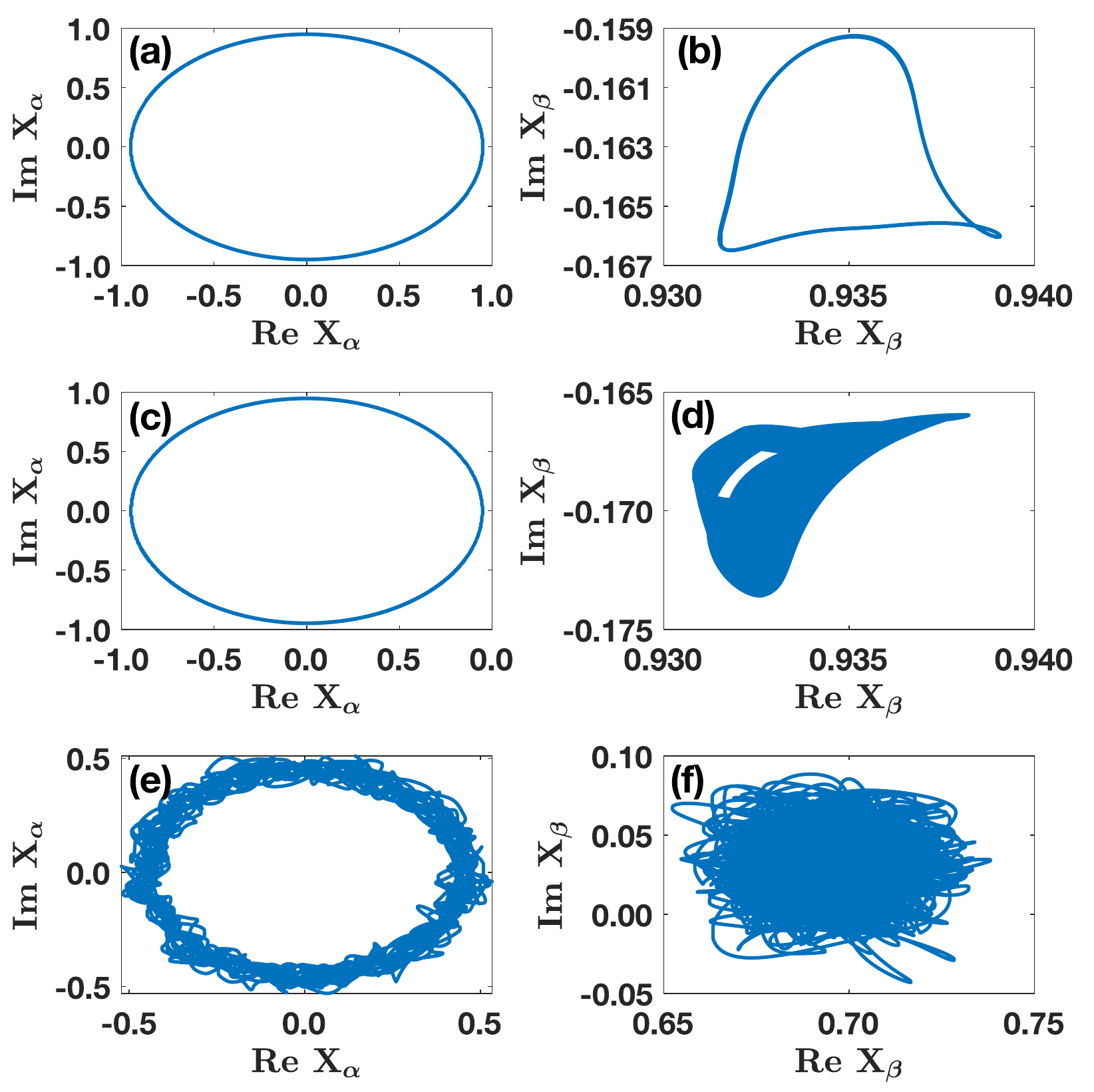}
\caption{{\bf Phase portraits of the order parameters.} The trajectories of the real and imaginary parts of the complex order  parameters $X_{\alpha}$ and $X_{\beta}$ are displayed for different values of the coupling constant $\sigma$ in the backward transition. These results are obtained by neglecting the transient,  in a network of $N=500$ nodes with $\sigma=3.4$ (panel (a) and (b)), with $\sigma=3.37$ (panel (c) and (d)) and  with $\sigma=1.69$ (panel (e) and (f)).}
\label{fig:5}       
\end{figure*}

\begin{figure*}[htbp!]
\centering
\includegraphics[width=1.4\columnwidth]{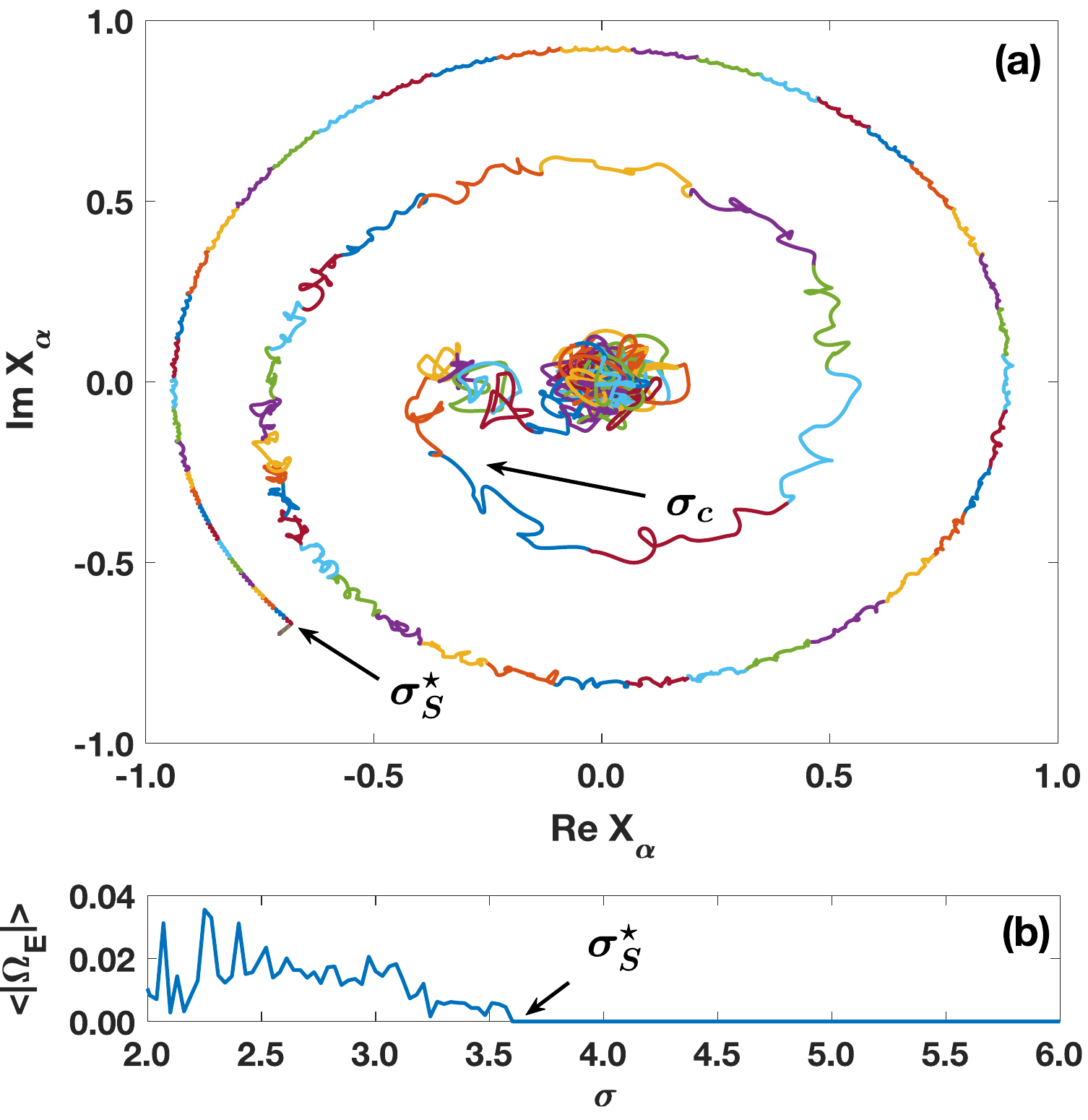}
\caption{ {{\bf The desynchronization transition.} The evolution of the complex order parameter $X_{\alpha}$ during the backward synchronization transition is plotted in panel (a) as the coupling constant decreases (each $\sigma$ step is indicated by the colors of different lines)}. These results have been obtained for a network of $N=500$, $T_{\text{max}}=10$, and  {$\Delta \sigma=0.03$}. During this backward transition, the onset of the rhythmic phase at $\sigma_S^{\star}$ and the onset of the incoherent phase at $\sigma_c$ are indicated.  The onset of the rhythmic phase corresponds exactly to the emergence of a non-zero oscillation frequency $\Omega_E$, as shown in panel (b).}
\label{fig:6}       
\end{figure*}
\subsection{Numerical investigation of the rhythmic phase}

In this section we investigate numerically the rhythmic phase, observed for $\sigma_c<\sigma<\sigma_S^{\star}$ in the backward transition, and for $\sigma_c^{\star}<\sigma<\sigma_S^{\star}$ in the forward transition.
In this region of the phase space, the system is in a non-stationary state where we can no longer assume that $X_{\alpha}$ and $X_{\beta}$ are stationary.

In order to study the dynamical behaviour of the complex order parameters characterized by slow fluctuations, we consider a fully connected network of size $N=500$, where we are able to follow the non-stationary dynamics for a long equilibration time $T_{\text{max}}$.

For $\sigma>\sigma_S^{\star}$, the order parameters are stationary as shown in Figure $\ref{fig:4}$(a) and Figure $\ref{fig:4}$(b). However, in  the  rhythmic phase, the order parameters do not reach a stable fixed point and  their real and imaginary parts undergo fluctuations as shown in Figure $\ref{fig:4}$(c)-(h).  In particular, close to the onset of the rhythmic phase $\sigma_S^{\star}$, the order parameter $X_{\alpha}$ displays a slow rotation in the complex plane with constant emergent frequency  $\Omega_E$, and constant absolute value $|X_{\alpha}|=R_{\alpha}$ as shown in Figure $\ref{fig:4}$(c). In this region, $X_{\beta}$  performs a  periodic motion along a closed limit cycle (see Figure $\ref{fig:4}$(d)). If the value of the coupling constant is decreased, first the order parameter $X_{\beta}$ displays a more complex dynamics (Figure $\ref{fig:4}$(f)) while $X_{\alpha}$ continues to oscillate at essentially constant absolute value $R_{\alpha}$ (Figure $\ref{fig:4}$(e)). As $\sigma$ approaches $\sigma_c$, higher frequency oscillations of the magnitude of $X_{\alpha}$ also set in, see Figure $\ref{fig:4}$(g). The  phase space portraits corresponding to the dynamics of the complex order parameters presented in Figure $\ref{fig:4}$ are shown in Figure $\ref{fig:5}$, revealing the nature of the fluctuations of the order parameters.

The dynamics of the complex order parameter $X_{\alpha}$ is particularly interesting in relation to the study of brain rhythms and cortical oscillations, which have their origin in the level of synchronization within neuronal populations or  cortical areas.
In order to describe the non-trivial dynamical behaviour of $X_{\alpha}$ during the backward transition, we show in Figure $\ref{fig:6}$ the phase portrait of $X_{\alpha}$
when the coupling constant $\sigma$ is decreased in time in a fully connected network of size $N=500,$ where at each value of the coupling constant the dynamics is equilibrated for a time  $T_{\text{max}}=10$. From Figure $\ref{fig:6}$(a), it is apparent that for $\sigma_c<\sigma<\sigma_S^{\star}$, the order parameter $X_{\alpha}$ displays slow frequency oscillations with an amplitude that decreases as the coupling constant $\sigma$ is decreased.  {Moreover, this complex time series reveals that the amplitude is also affected on very short time scales by fluctuations of small amplitude and much faster frequencies. These are seen to become increasingly significant as the coupling constant approaches $\sigma_c$}.  {In Figure $\ref{fig:6}$(b), we highlight the region close to the transition between stationarity and rhythmic phase}. We confirm excellent agreement with the theoretical estimate of the onset of the steady state (see Eq. ($\ref{sigma_s_est}$) in Section $\ref{sec:stationaryphases}$) and compare this to the measured frequency of oscillation $\Omega_E$. As expected, the onset of non-stationarity coincides with the emergence of oscillations, as seen in the phase space evolution and direct measurement of $\Omega_E$.

\begin{figure*}[htbp!]
\centering
\includegraphics[width=1\columnwidth]{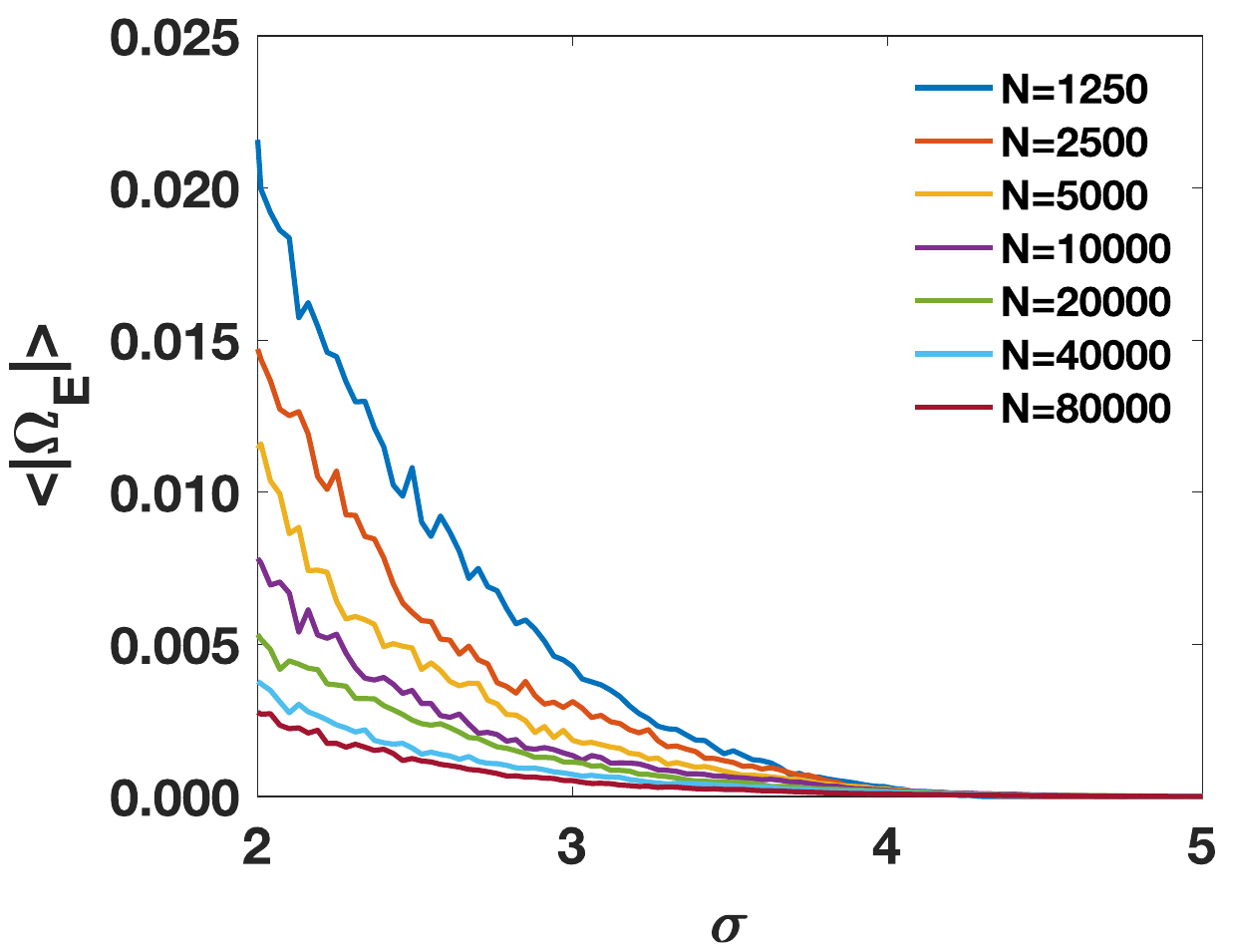}
\caption{{\bf The emergent frequency $\Omega_E$ characterizing the rhythmic phase.} The absolute value of the emergent frequency  $|\Omega_E|$ characterizing  the rhythmic phase is shown as a function of $\sigma$, for networks of different network sizes $N$. The data is averaged  over $100$ realizations of the intrinsic frequencies along the forward transition. The equilibration time taken is $T_{\text{max}} = 10$.}
\label{fig:8}       
\end{figure*}

Finally, we numerically observe that the angular frequency of oscillations of $X_\alpha$ in the rhythmic phase depends on the network size, as well as the coupling strength. In order to evaluate this effect, we have conducted simulations of Dirac synchronization with varying system size. For each network size, we performed $100$ independent iterations of the phase diagram. We show in Figure $\ref{fig:8}$ the average of the oscillation frequency $\Omega_E$ measured at each $\sigma$ step in the rhythmic phase along the forward transition for varying $N$. These simulations were obtained with equilibration time $T_{\text{max}} = 10$, and the individual frequencies measured for each iteration are averaged over the last fifth of the time series.

These results confirm that the oscillation frequency of the order parameter $X_\alpha$ decreases with stronger coupling strength and reaches $0$ at the onset of steady state for all finite system sizes. These extended numerical simulations also confirm that this corresponds precisely to the predicted onset of the steady state discussed in Section $\ref{sec:stationaryphases}$, as shown for a single iteration with $N = 500$ in Figure $\ref{fig:6}$. As expected, we also confirm that the rhythmic phase extends further for larger systems. Moreover, Figure $\ref{fig:8}$ clearly reveals that the oscillation frequency of the coherent phase decreases as the systems considered increase in size. 

\subsection{Theoretical treatment of  Dirac synchronization}
\label{theory}
\subsubsection{The continuity equation}
 {We can obtain analytical stationary solutions to the dynamics of Dirac synchronization in Eqs. (\ref{angledot}) by using a continuity equation approach \cite{Strogatz_Mirollo}. This approach is meant to capture the dynamics of the distribution of the phases $\alpha_i$ and $\beta_i$.}
To this end let us define the density distribution $\rho^{(i)}(\alpha,\beta|\omega_i,\hat{\omega}_i)$ of the phases $\alpha_i$ and $\beta_i$ given the frequencies $\omega_i$ and $\hat{\omega}_i$.  {Since the phases obey the deterministic Eq.(\ref{angledot}), it follows that the time evolution of this density distribution is dictated by the continuity equation} 
\bea
&&\frac{\partial \rho^{(i)}(\alpha,\beta|\omega_i,\hat{\omega}_i)}{\partial t}+\nabla\cdot {\bf J}_i=0,
\label{continuity}
\eea
where the current ${\bf J}_i$ is defined as 
\bea
{\bf J}_i= \rho^{(i)}(\alpha,\beta|\omega_i,\hat{\omega}_i){\bf v}_i
\eea
and $\nabla=(\partial_{\alpha},\partial_{\beta})$.
Here the velocity vector ${\bf v}_i$ is given by 
\bea
{\bf v}_i={\bm \kappa}_i+\sigma \mbox{Im}\left[{\bf \hat X}e^{-\textrm{i}{\bm\alpha}_i}\right].
\eea

In order to solve the continuity equation, we extend the Ott-Antonsen \cite{Ott_Antonsen} approach to this 2-dimensional case, making the ansatz that the Fourier expansion of $\rho^{(i)}(\alpha,\beta|\omega_i,\hat{\omega}_i)$ can be expressed as
\begin{widetext}
\bea
\rho^{(i)}(\alpha,\beta|\omega,\hat{\omega})=\frac{1}{(2\pi)^2}\left\{1+\sum_{n>0}\left[f_{n}^{(\alpha,i)}(\omega,\hat{\omega},t) e^{\textrm{i}n\alpha}+c.c.\right]\right\}\left\{1+\sum_{m>0}\left[f_{m}^{(\beta,i)}(\omega,\hat{\omega},t) e^{\textrm{i}m\beta}+c.c.\right]\right\},
\label{ott}
\eea
\end{widetext}
 {where $c.c.$ denotes the complex conjugate of the preceding quantity. The expansion coefficients are given by} 
\bea
f_{n}^{(\alpha,i)}(\omega,\hat{\omega},t)=[a_i(\omega,\hat{\omega},t)]^n\nonumber \\
f_{m}^{(\beta,i)}(\omega,\hat{\omega},t)=[b_i(\omega,\hat{\omega},t)]^m
\label{two}
\eea
for $n>0,m>0$. The series in Eq. (\ref{ott}) converges for $|a_i|\leq 1$ and $|b_i|\leq 1$ provided that we attribute to $\alpha$  {and $\beta$} an infinitesimally small imaginary part.

For $X_{\alpha}\neq 0$, $a_i\neq 0 $ and $b_i\neq 0$, the continuity equation is satisfied if and only if (see Appendix \ref{Ap1} for details) $a_i$ and $b_i$ are complex variables with absolute value one, i.e. $|a_i|=|b_i|=1$, that satisfy the system of differential equations 
\begin{widetext}
 {\bea
\partial_t a_i+\textrm{i}a_i\kappa_{\alpha,i}
+\frac{1}{2}\sigma \left[X_{\alpha}a^{2}_i-X_{\alpha}^{\star}\right]
-\sigma \frac{1}{4}a_i(b_i-b_i^{-1})=0, \nonumber \\
\partial_t b_i+\textrm{i}b_i\kappa_{\beta,i}
+\frac{1}{2}\sigma \bar{c}\left[X_{\alpha}a_i-X_{\alpha}^{\star}a_i^{-1}\right]b_i
+\sigma \frac{1}{2}(b_i^{2}-1)=0,
\label{dyn2Ma}
\eea}
\end{widetext}
where here and in the following we indicate with $X_{\alpha}^{\star}$ the complex conjugate of $X_{\alpha}$, and with $a_i^{-1}=a_i^{\star}$ and $b_i^{-1}=b_i^{\star}$ the complex conjugate of $a_i$ and $b_i$ respectively.
We note that the only stationary solutions of these equations are 
\bea
a_i&=&-\textrm{i}d_{i,\alpha}\pm \sqrt{1-d_{i,\alpha}^2}\nonumber \\
b_i&=&-\textrm{i}d_{i,\beta}\pm \sqrt{1-d_{i,\beta}^2},\label{stat_b}
\eea
with  $d_{i,\alpha}$, $d_{i,\beta}$ defined as 
\bea
d_{i,\alpha}&=&\frac{\omega_i-\hat{\Omega}}{\sigma  R_{\alpha}}, \nonumber \\
d_{i,\beta}&=&-{\hat{\omega}_i}/{\sigma}+\textrm{Im} X_{\beta},
\label{dibeta}
\eea
and having absolute value $|d_{i,\alpha}|\leq 1$ and $|d_{i,\beta}|\leq 1$.
This last constraint is necessary to ensure $|a_i|=|b_i|=1$.
This result is very different from the corresponding result for the standard Kuramoto model because it implies that the coherent phase with $R_{\alpha}>0$ cannot be a stationary solution in the infinite network limit as long as $\bm{\omega}$ and $\hat{\bm{\omega}}$ are drawn from an unbounded distribution. Indeed, a necessary condition to have all the phases in a stationary state implies that $|d_{i,\alpha}|\leq 1$ and $|d_{i,\beta}|\leq 1$ for every node of the network. This implies in turn that the frequencies $\bm \omega$ and $\bm{\hat\omega}$ are bounded as we will discuss in the next sections.
 
In the case in which $X_{\alpha}=0$, instead, the continuity equation is satisfied if and only if (see Appendix \ref{Ap1} for details) $a_i$ and $b_i,$ with $|a_i|\neq 0$ and $|b_i|=1,$ follow the system of differential equations 
 {\bea
&&\partial_t a_i
+\textrm{i}a_i\kappa_{\alpha,i}
-\sigma \frac{1}{4}a_i(b_i-b_i^{-1})=0\nonumber\\
&&\partial_t b_i+\textrm{i}b_i\kappa_{\beta,i}
+\sigma \frac{1}{2}(b_i^{2}-1)=0.
\label{dyn2M}
\eea}
For $|a_i|=0$, the equation for $b_i$ is unchanged, but $b_i$ can have an arbitrary large absolute value.
In this last case, the steady state solution of  Eqs.~(\ref{dyn2M})  reads
\bea
a_i&=&0\label{stat_aa2} \nonumber \\
b_i&=&-\textrm{i}\hat{d}_{i,\beta}\pm \sqrt{1-\hat{d}_{i,\beta}^2},\label{stat_b2}
\eea
with 
\bea
\hat{d}_{i,\beta}&=&\kappa_{\beta,i}/{\sigma}.
\label{dibeta2}
\eea

Let us now derive the equations that determine the order parameters $X_{\alpha}$ and $X_{\beta}$ for any possible value of the coupling constant $\sigma$.
Let us assume that the frequencies $\omega_i$ and $\hat{\omega}_i$ associated to each node $i$ are known. With this hypothesis, the complex order parameters can be expressed in terms of the density $\rho^{(i)}(\alpha,\beta|\omega_i,\hat\omega_i)$ as
\bea
X_{\alpha}&=&\frac{1}{N}\sum_{i=1}^N\int d\alpha\int d\beta\ \rho^{(i)}(\alpha,\beta|\omega_i,\hat\omega_i)e^{\textrm{i}\alpha},\nonumber \\
X_{\beta}&=&\frac{1}{N}\sum_{i=1}^N\int d\alpha\int d\beta\ \rho^{(i)}(\alpha,\beta|\omega_i,\hat\omega_i)e^{\textrm{i}\beta}.\label{eq:Xab}
\eea

When $\rho^{(i)}(\alpha,\beta|\omega,\hat{\omega})$ satisfies the generalized Ott-Antonsen ansatz given by Eq. (\ref{ott}) and Eq. $(\ref{two})$, these complex order parameters can be expressed in terms of the functions $a_i(\omega_i,\hat{\omega}_i)$ and $b_i(\omega_i,\hat{\omega}_i)$ as
\bea
X_{\alpha}&=&\frac{1}{N}\sum_{i=1}^Na_i^{\star}(\omega_i,\hat{\omega}_i),\nonumber  \\
X_{\beta}&=&\frac{1}{N}\sum_{i=1}^N b_i^{\star}(\omega_i,\hat{\omega}_i),\label{defXb}
\eea
where $a_i^{\star}$ and $b_i^{\star}$ are the complex conjugates of $a_i$ and $b_i$, respectively.

If the internal frequencies $\omega_i$ and $\hat{\omega}_i$ are not known, we can express these complex order parameters in terms of the marginal distributions $G_0(\omega)$ and $G_1(\hat{\omega})$ as 
\bea
X_{\alpha}&=&\int d\omega \int d\hat{\omega}G_0(\omega)G_1(\hat{\omega})a^{\star}(\omega,\hat{\omega}),\nonumber \\
X_{\beta}&=&\int d\omega \int d\hat{\omega}G_0(\omega)G_1(\hat{\omega}) b^{\star}(\omega,\hat{\omega}).\label{defXbG}
\eea
 {This derivation shows that $a(\omega,\hat{\omega})$ and $b(\omega,\hat{\omega})$ can be obtained from the integration of (Eqs. \ref{dyn2Ma} and \ref{dyn2M}). In particular, as we discuss in the next paragraph (paragraph \ref{sec:stationaryphases}) these equations will be used  to  investigate the steady state solution of this dynamics and the range of frequencies on which this stationary solution can be  observed. However for Dirac synchronization we observe a phenomenon that does not have an equivalent in the standard Kuramoto model. Indeed Eqs. \ref{dyn2Ma} and \ref{dyn2M} do not always admit a coherent stationary solution, and actually the non-stationary phase, also called rhythmic phase, is the stable one in the large network limit. In this case we observe that  Eqs. \ref{dyn2Ma} and \ref{dyn2M} are equivalent to Eqs. (\ref{angledot}) and therefore even their numerical integration is not advantageous with respect to the numerical integration of the original dynamics. Therefore to capture analytically the phase diagram of Dirac synchronization in the  rhythmic phase  we rely on approximate  derivations which we cover in paragraph \ref{sec:Theo_Rhy_Phase}}.

\subsubsection{The stationary phases of Dirac synchronization}
\label{sec:stationaryphases}
In the previous section we have shown that Dirac synchronization admits a stationary state in the following two scenarios:
\begin{itemize}
\item[-] {\em Incoherent Phase -} This is  the incoherent phase where for each node $i$, $a_i$ and $b_i$ are given by Eqs. (\ref{stat_b2}).
In this phase, $R_{\alpha}=0$ and the order parameter  {$X_{\beta}=R_{\beta}e^{i\eta_{\beta}}$}  is determined by the equations
 {\bea
{R}_{\beta}\cos\eta_{\beta}&=&\frac{1}{N}\sum_{i=1}^N\sqrt{1-\hat{d}_{i,\beta}^2}H(1-\hat{d}_{i,\beta}^2),\nonumber \\
{R}_\beta\sin\eta_{\beta}&=&\frac{1}{N}\sum_{i=1}^N\hat{d}_{i,\beta}H(1-\hat{d}_{i,\beta}^2),
\label{comp}
\eea}
where  $\hat{d}_{i,\beta}$ is given by Eq.~(\ref{dibeta2}) and $H(\cdot)$ is the Heaviside step function. 
Finally, if the intrinsic frequencies $\bm\omega$ and projected frequencies $\hat{\bm{\omega}}$ are not known, we can average $a_i^{\star}(\omega,\hat{\omega})$ and $b_i^{\star}(\omega,\hat{\omega})$ appearing in Eqs.~(\ref{defXb}) over the  marginal distributions $G_0(\omega)$ and $G_1(\hat{\omega})$.
We observe that the steady state  Eqs.~(\ref{comp}) always has a solution compatible with  {$\eta_{\beta}=0$}, indicating that the  contribution from the phases ${\beta}_i$ that are drifting is null. Therefore, $R_{\beta}$ in the incoherent  phase is given by 
\bea
\hspace*{-10mm}{R}_\beta&=&\int d\omega\int_{|\hat{d}_{i,\beta}|\leq1} d\hat{\omega} G_0({\omega}) G_1(\hat{\omega})\sqrt{1-\hat{d}_{i,\beta}^2}.
\label{comp2a}
\eea
\item[-]{\em Stationary Coherent Phase -}  This is the coherent phase where for each node $i$, $a_i$ and $b_i$ are given by Eqs.~(\ref{stat_b}), respectively.
Note that this phase differs significantly from the coherent phase of the standard Kuramoto model where drifting phases can also give rise to a stationary continuity equation. Indeed the constraints $|a_i|=|b_i|=1$ imply that this phase can only be observed when there are no drifting phases, and for each node $i$ the phases $\alpha_i,\beta_i$ are frozen.  
This can only occur in finite size networks, provided that the coupling constant $\sigma$ is sufficiently large. Indeed, $|a_i|=|b_i|=1$ implies that $|d_{i,\alpha}|\leq 1$ and $|d_{i,\beta}|\leq 1$ for all nodes $i$ of the network.
By using the explicit expression of $d_{i,\alpha}$ and $d_{i,\beta}$ given by the Eqs.~$(\ref{dibeta})$, this implies that the stationary coherent phase can only be achieved if 
\bea
\sigma\geq \frac{1}{R_{\alpha}}\max_{i}|\omega_i-\hat{\Omega}|,\nonumber \\
 {\sigma\geq \max_{i}|\hat{\omega}_i-\mbox{Im}X_{\beta}|,}
\label{sigmau}
\eea
are simultaneously satisfied, which gives 
\begin{align}
\sigma_S^{\star} = \max\left(\max_i\frac{|\omega_i-\hat{\Omega}|}{R_{\alpha}}, {\max_i|\hat{\omega}_i-\mbox{Im}X_{\beta}|}\right).
\label{sigma_s_est}
\end{align}

We numerically find excellent agreement with this estimate, as discussed previously.

Using a similar derivation as the one outlined for the incoherent phase, we can derive the expression for $R_{\alpha}$ and $R_{\beta}$ in the stationary coherent phase, which can be expressed as 
\bea
R_{\alpha}&=&\frac{1}{N} \sum_{i=1}^N\sqrt{1-d_{i,\alpha}^2},\nonumber \\
{R}_\beta&=&\frac{1}{N}
\sum_{i=1}^N \sqrt{1-d_{i,\beta}^2},
\label{comp2}
\eea
where $d_{i,\alpha}$ $d_{i,\beta}$ are both smaller than one in absolute value and given by Eqs.~(\ref{dibeta}).
If the frequencies of the individual nodes are not known,  {averaging over the distributions $G_0(\omega)$ and $G_1(\hat{\omega})$} one finds that $\mbox{Im}X_{\beta}=0$ and  the order parameters $R_{\alpha}$ and $R_{\beta}$ can be expressed as 
\bea
R_\alpha&=& \int d\omega G_0(\omega)\sqrt{1-d_{i,\alpha}^2},\nonumber \\
R_\beta&=&\int  d\hat{\omega}G_1(\hat{\omega})\sqrt{1-d_{i,\beta}^2}.
\eea
From this discussion, and since this phase can only be observed for a coupling constant $\sigma$ satisfying Eqs.~(\ref{sigmau}), it follows that this stationary coherent phase can only be achieved if the internal frequencies $\bm\omega$ and $\hat{\bm{\omega}}$ are bounded. Since the internal frequencies are Gaussian distributed, this implies that the stationary coherent phase is only observed in finite size networks at a value of the coupling constant that increases with the network size $N$. We were able to verify this effect numerically, finding that $\sigma_S^\star$ clearly increases with $N$.
\end{itemize}

\subsubsection{The theoretical interpretation of the non-stationary rhythmic phase}
\label{sec:Theo_Rhy_Phase}
From the  theoretical treatment of the stationary phases of Dirac synchronization  performed in the previous paragraph, we draw the important conclusion that the coherent phase of Dirac synchronization is non-stationary in the thermodynamic limit. Indeed, in this phase, the continuity equation is characterized by a non-vanishing current. This is very different phenomenology in comparison with the standard Kuramoto model, where the drifting phases can still coexist with a stationary continuity equation.\\

In order to treat the coherent non-stationary phase (the rhythmic phase), we provide here an approximate theoretical framework that captures the essential physics of this dynamics.
Our starting point is the dynamical equation (Eq. (\ref{angledot})) obeyed by the vector $\bm\alpha_i=(\alpha_i,\beta_i)^{\top}$ of phases associated to each node $i$. We also make the important numerical observation that in the non-stationary coherent phase, the order parameter  {$X_{\alpha}=R_{\alpha}e^{\textrm{i}\eta_{\alpha}}$} acquires a phase velocity even if $\hat{\Omega}=0$. We characterize this phase as a rhythmic phase with  {\bea
\eta_{\alpha}(t)\simeq\eta_{\alpha}(0)+{\Omega_E}t,
\label{phi_alpha}
\eea}
where we note, however, that numerical simulations reveal that the observed emergent frequency ${\Omega_E}$ decreases with increasing network size (see Figure $\ref{fig:8}$).

While in the standard Kuramoto model the phases are either frozen in the rotating moving frame of the order parameter or drifting, in Dirac synchronization the scenario is richer, because some phases are allowed to oscillate but still contribute to the order parameters. 
Indeed, by studying Eq.~(\ref{angledot}) with the hypothesis that  {$\eta_{\alpha}$} obeys Eq.~(\ref{phi_alpha}), we can classify the phases $\bm\alpha_i$ associated to each node $i$ of the network into four classes (whose typical trajectories are shown in Figure \ref{fig:classification}):

\begin{figure}[tbh]
\centering
\includegraphics[width=1.0\columnwidth]{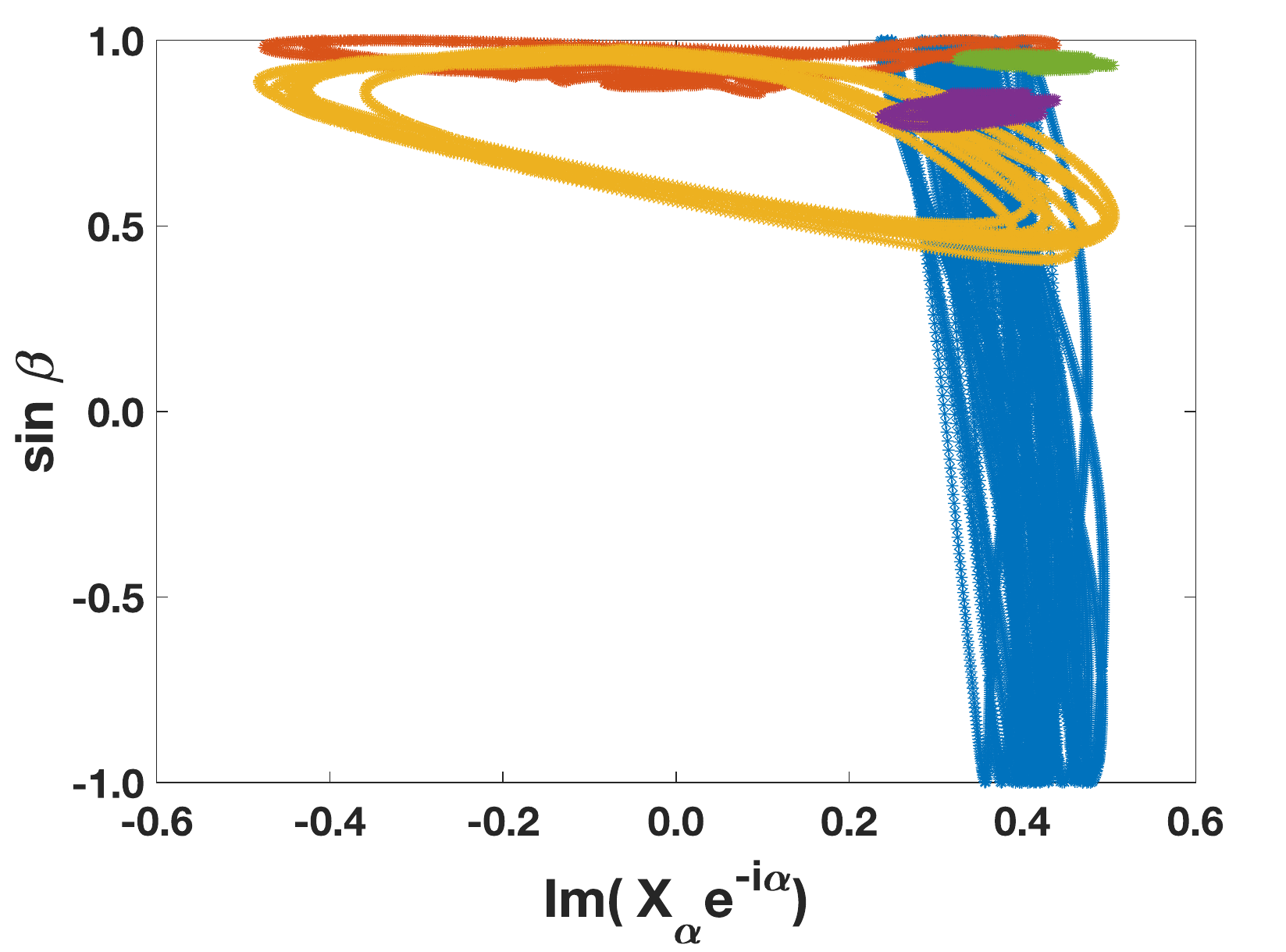}
\caption{ {\bf Classification of nodes based on the trajectory of their phases.} The trajectory of the $(\alpha_i,\beta_i)$ phases is shown in the plane $(\mbox{Im}(X_{\alpha}e^{-\textrm{i}\alpha}),\sin\beta)$ for nodes with frozen phases (green and purple trajectories), for nodes with $\alpha$-oscillating phases (blue trajectory) and nodes with $\beta$ oscillating phases (red and yellow trajectories). Data are obtained from numerical simulation of Dirac synchronization in their backward transition for a value of the coupling constant $\sigma=1.77$ and network size $N=500$.}
\label{fig:classification}       
\end{figure}

\begin{figure*}[tbh]
\centering
\includegraphics[width=1.8\columnwidth]{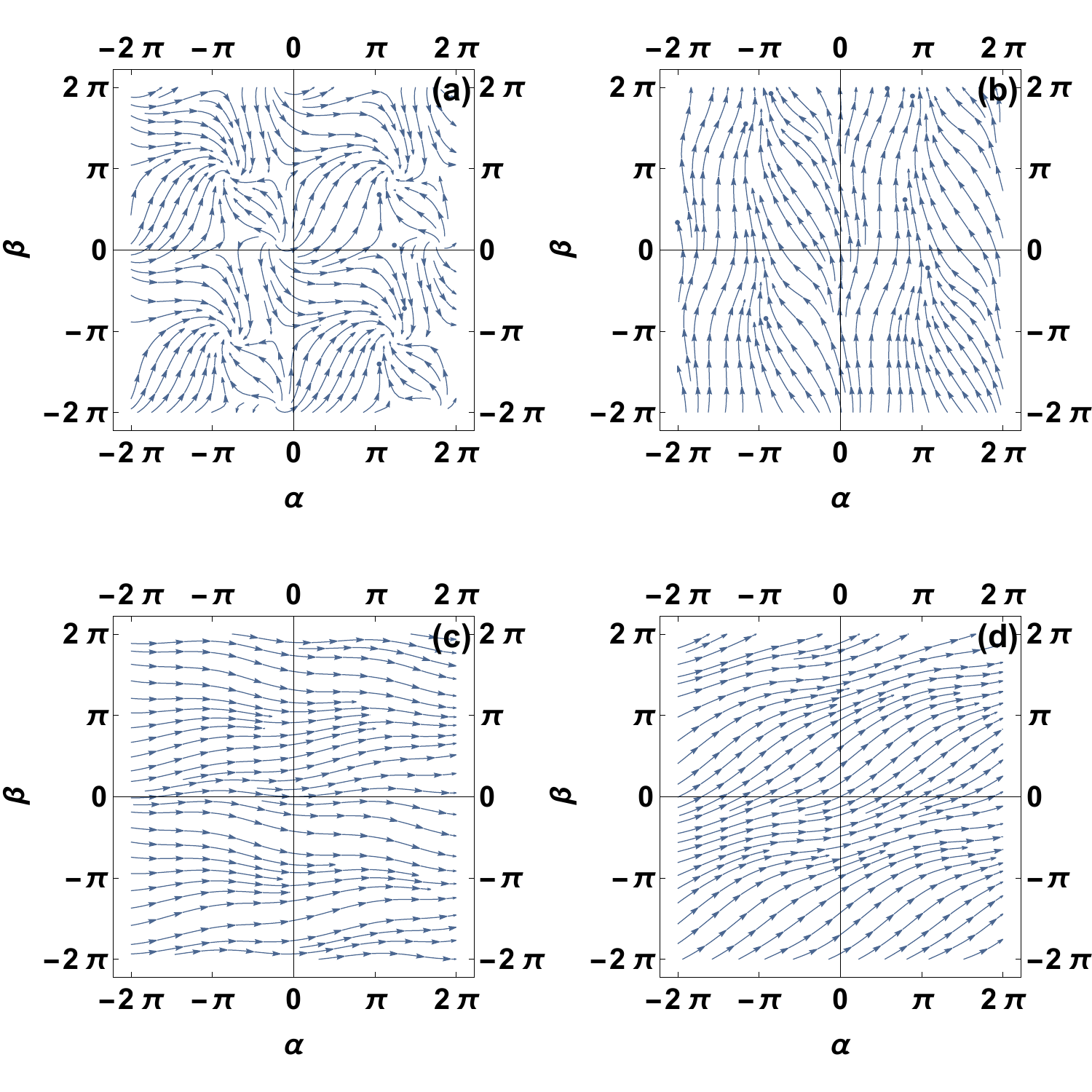}
\caption{{\bf Streamplots of the dynamical equations of Dirac synchronization.}  Streamplots of  Eq.(\ref{angledot}), for $X_{\alpha}=R_{\alpha}=0.8$ ($\tilde{\Omega}=0$) and $\sigma=2$. The four different streamplots corresponds to nodes with frozen phases (panel a) $\omega=1$, $\hat{\omega}=0.5$, to nodes with $\alpha$-oscillating phases (panel b) $\omega=1$, $\hat{\omega}=-3.5$, to nodes with $\beta$-oscillating phases (panel c) $\omega=8$, $\hat{\omega}=8$, to nodes with drifting phases (panel d) $\omega=7$, $\hat{\omega}=3$.}
\label{fig:streamplot}       
\end{figure*}

\begin{itemize}
\item[(a)] {\em Nodes with frozen phases-} These are nodes $i$ that for large times have both $\alpha_i$ and $\beta_i$ phases frozen in the rotating frame.  A typical streamplot of the phases of these nodes is represented in Figure $\ref{fig:streamplot}$(a).
Under the simplifying assumption that only the order parameter $X_{\alpha}$ rotates with frequency ${\Omega_E}$, these phases obey 
\bea
\dot{\alpha}_i={\Omega_E},\quad
\dot{\beta}_i=0.
\eea
Therefore, in the rotating frame of the order parameters these phases are frozen on the values 
\bea
 {\sin(\alpha_i-\eta_{\alpha})}= {d}_{i,\alpha},\nonumber \\
\sin \beta_i= d_{i,\beta}.
\label{eq:f}
\eea
with 
\bea
d_{i,\alpha}&=&\frac{\omega_i-\hat{\Omega}-{\Omega_E}/(1+\bar{c}/2)}{\sigma R_{\alpha}}\nonumber\\
d_{i,\beta}&=&\frac{-\hat{\omega}_i+\bar{c}{\Omega_E}/(1+\bar{c}/2)}{\sigma}+\mbox{Im}X_{\beta}.
\eea
A necessary condition for the nodes to be frozen is that  
\bea
|d_{i,\alpha}|\leq 1,\quad |d_{i,\beta}|\leq 1,
\eea
so that Eqs.~(\ref{eq:f}) are well defined. Therefore in the plane $(\mbox{Im}(X_{\alpha}e^{-\textrm{i}\alpha}),\sin\beta)$, these phases in theory should appear as a single dot. Since in the real simulations the order parameter $X_{\beta}$ has a non-zero phase, these phases appear in Figure $\ref{fig:classification}$ as little localized clouds instead of single dots.
\item[(b)]{\em Nodes with $\alpha$-oscillating phases-} These are nodes $i$ that for large times have the phase $\beta_i$ drifting, while the phase $\alpha_i$ oscillates in a non-trivial way contributing to the order parameter $X_{\alpha}$. A typical streamplot of the phases of these nodes is represented in Figure $\ref{fig:streamplot}$(b). Asymptotically in time, these phases obey
\bea
\dot{\alpha}_i={\Omega_E},\quad
\overline{\sin {\beta}_i}\simeq 0,
\eea
where we denote a late-time average with an overbar. 
By inserting these conditions in the dynamical Eq.~(\ref{angledot}), it follows that  
\bea
\overline{\mbox{Im}(X_{\alpha}e^{-\textrm{i}\alpha_i}})\simeq &\frac{1}{\sigma}({\Omega_E}-\kappa_{\alpha,i}).
\eea
Approximating  {$\overline{\mbox{Im}(X_{\alpha}e^{-\textrm{i}\alpha_i})}\simeq R_{\alpha} \sin \overline{(\eta_{\alpha}-\alpha_i)}$}, we get 
 {\bea
\sin\overline{(\alpha_i-\eta_{\alpha})}\simeq  \hat{d}_{i,\alpha}\equiv\frac{1}{\sigma R_{\alpha}}(\kappa_{\alpha,i}-\Omega_E).
\eea}
From these arguments, it follows that these $\alpha_i$ oscillating phases are encountered when  
\bea
|\hat{d}_{i,\alpha}|\leq 1,\quad
|d_{i,\beta}|\geq 1.
\eea
In the plane $(\mbox{Im}(X_{\alpha}e^{-\textrm{i}\alpha}),\sin\beta)$, these phases have a trajectory that spans only a limited range of the values of $\mbox{Im}(X_{\alpha}e^{-\textrm{i}\alpha})$ while the $\beta_i$ phases are drifting (see the blue trajectory in Figure $\ref{fig:classification}$).
\item[(c)]{\em Nodes with $\beta$-oscillating phases-}
These are nodes $i$ that at large times have the phase $\alpha_i$ drifting, while the phase $\beta_i$ oscillates in a non-trivial way, contributing to the order parameter $X_{\beta}$.  A typical streamplot of the phases of these nodes is represented in Figure $\ref{fig:streamplot}$(c). Therefore, asymptotically in time,  the phases of these nodes obey
\bea
\overline{\mbox{Im}(X_{\alpha}e^{-\textrm{i}\alpha_i}})\simeq 0,\quad
{\dot{\beta}_i}=0.
\eea
By inserting these conditions in the dynamical Eq.~(\ref{angledot}), we conclude that 
\bea
\overline{\sin \beta_i}\simeq  -\hat{d}_{i,\beta} \equiv \frac{-\kappa_{\beta,i}}{\sigma}.
\eea
By approximating $\overline{\sin\beta_i}\simeq \sin \overline \beta_i$, we obtain the following estimation of $\sin\overline{\beta_i}$ 
\bea
\sin\overline{\beta_i}\simeq - \hat{d}_{i,\beta}=\frac{-\kappa_{\beta,i}}{\sigma}.
\eea
It follows that these nodes are encountered when the following conditions are satisfied:
\bea
|\hat{d}_{i,\beta}|\leq 1,\quad
|d_{i,\alpha}|\geq 1.
\eea
In the plane $(\mbox{Im}(X_{\alpha}e^{-\textrm{i}\alpha}),\sin\beta)$, these phases have a trajectory that spans only a limited range of the values of  $\sin \beta_i$ while the $\alpha_i$ phases are drifting (see the red and yellow trajectories in Figure $\ref{fig:classification}$). 
\item[(d)]{\em Nodes with drifting phases-}
These are nodes whose  phases do not satisfy any of the previous conditions, where one observes 
\bea
\overline{\mbox{Im}(X_{\alpha}e^{-\textrm{i}\alpha_i}})\simeq 0,\quad
\overline{\sin\beta_i}\simeq 0.
\eea
A typical streamplot of the phases of these nodes is represented in Figure $\ref{fig:streamplot}$(d).
The phases of these nodes do not contribute to any of the order parameters.
\end{itemize}
The frozen phases and the $\alpha$-oscillating phases both contribute to the order parameter $R_{\alpha}$ while the frozen phases and the $\beta$-oscillating phases both contribute to $R_{\beta}$. Therefore, the order parameters $R_{\alpha}$ and $R_{\beta}$ can be approximated by 
\bea
R_{\alpha}&\simeq&\frac{1}{N}\sum_{i=1}^N\sqrt{1-d_{i,\alpha}^2}H(1-d_{i,\alpha}^2)H(1-d_{i,\beta}^2)\nonumber \\
&&+ \frac{1}{N} \sum_{i=1}^N\sqrt{1-\hat{d}_{i,\alpha}^2}H(1-\hat{d}_{i,\alpha}^2)H(d_{i,\beta}^2-1),\nonumber \\
{R}_\beta &\simeq &\frac{1}{N}
\sum_{i=1}^N \sqrt{1-d_{i,\beta}^2}H(1-d_{i,\alpha}^2)H(1-d_{i,\beta}^2)\nonumber \\
&&+ \frac{1}{N} \sum_{i=1}^N\sqrt{1-\hat{d}_{i,\beta}^2}H({d}_{i,\alpha}^2-1)H(1-\hat{d}_{i,\beta}^2).
\label{comp2ns}
\eea
where $d_{i,\alpha}$, $d_{i,\beta}$ are given by Eq. (\ref{dibeta}) and $\hat{d}_{i,\alpha}$, $\hat{d}_{i,\beta}$ are given by 
\bea
\hat{d}_{i,\alpha}&=&\frac{\kappa_{\alpha,i}-{\Omega_E}}{\sigma R_{\alpha}},\nonumber\\
\hat{d}_{i,\beta}&=&\frac{\kappa_{\beta,i}}{\sigma}.
\eea
If the frequencies of the individual nodes are not known, in the approximation in which $\mbox{Im}X_{\beta}\simeq 0$, the order parameters $R_{\alpha}$ and $R_{\beta}$ can be expressed as 
\begin{widetext}
\bea
R_\alpha &\simeq &\int_{|d_{i,\alpha}|\leq 1} d\omega G_0(\omega)\int_{|d_{i,\beta}|\leq 1}   d\hat{\omega}G_1(\hat{\omega})\sqrt{1-d_{i,\alpha}^2}+\int_{|\hat{d}_{i,\alpha}|\leq 1} d\omega G_0(\omega)\int_{|d_{i,\beta}|\geq 1}   d\hat{\omega}G_1(\hat{\omega})\sqrt{1-\hat{d}_{i,\alpha}^2},\nonumber\\
R_\beta&\simeq&\int_{|d_{i,\alpha}|\leq 1} d\omega G_0(\omega)\int_{|d_{i,\beta}|\leq 1}   d\hat{\omega}G_1(\hat{\omega})\sqrt{1-d_{i,\beta}^2}\int_{|d_{i,\alpha}|\geq 1} d\omega G_0(\omega)\int_{|\hat{d}_{i,\beta}|\leq 1}   d\hat{\omega}G_1(\hat{\omega})\sqrt{1-\hat{d}_{i,\beta}^2}.
\eea
\end{widetext}
We now make use of the numerical observation that $\Omega_E$ decreases with the network size, implying that the period of the oscillations of the order parameter $X_{\alpha}$ becomes increasingly long with increasing network sizes.
By substituting $\Omega_E=0$, these equations can be used to determine the order parameters $R_{\alpha}$ and $R_{\beta}$ as a function of the coupling constant $\sigma$ in the limit $N\to \infty$. Indeed, these are the equations that provide the theoretical expectation of the phase diagram in Figure $\ref{fig:2}$.
Moreover, these equations can be used to estimate the critical value $\sigma_c$ by substituting $\Omega_E=0$ and expanding the self-consistent expression of $R_{\alpha}$ for small values of $R_{\alpha}$. To this end, we write the self-consistent equation for $R_{\alpha}$ as 
\begin{widetext}
\bea
\hspace{-5mm}1&=&\sigma\int_{-1}^{1}dx G_{0}(\Omega_0+\sigma R_{\alpha}x)\sqrt{1-x^2}\int_{-\sigma}^{\sigma}d\hat{\omega}G_1(\hat{\omega})+\sigma\int_{|\hat{\omega}|\geq \sigma}d\hat{\omega}G_1(\hat{\omega})\int_{-1}^{1}dx G_{0}(\Omega_0+\sigma R_{\alpha}x-\hat{\omega}/2)\sqrt{1-x^2}. 
\eea
\end{widetext}
For $R_{\alpha}\ll 1$, we make the following approximations
\bea
G_{0}(\Omega_0+x R_{\alpha}x)&\simeq & G_0(\Omega_0),\nonumber \\
G_{0}(\Omega_0+\sigma R_{\alpha}x-\hat{\omega}/2)&\simeq& G_0(\Omega_0-\hat{\omega}/2).
\eea
Inserting these expressions into the self-consistent equation for $R_{\alpha}$, we can derive the equation determining the value of the coupling constant $\sigma=\sigma_c$ at which we observe the continuous phase transition,
\bea
1=\sigma \sqrt{\frac{\pi}{2}}\left[\frac{1}{2}\mbox{erf}\left(\frac{\sigma}{\sqrt{2}}\right)+\frac{1}{\sqrt{5}} \mbox{erfc}\left(\frac{1}{2}\sqrt{\frac{5}{2}}\sigma\right)\right]
\eea
 {where $\mbox{erf}(x)$ is the error function and $\mbox{erfc}(x)$ is the complementary error function.} 
This equation can be solved numerically, providing the value of $\sigma_c$ given by
\bea
\sigma_c=1.66229\ldots.
\eea
Similarly, we can consider the classification of the different phases of oscillators to study the onset of the instability of the incoherent phase. We obtain the estimate for $\sigma_c^{\star}$ given by (see Appendix \ref{crit} for details) 
\bea
\sigma_c^{\star}=2.14623\ldots.
\eea

\section{Conclusions}
In this work, we have formulated and discussed the explosive Dirac synchronization of locally coupled topological signals associated to the nodes and to the links of a network. Topological signals associated to nodes are traditionally studied in models of non-linear dynamics of a network. However, the dynamics of topological signals associated to the links of the network is so far much less explored.

In brain and neuronal networks, the topological signals of the links can be associated to a dynamical state of synapses (for example oscillatory signals associated to intracellular calcium dynamics involved in synaptic communication among neurons \cite{astrocyte1997}), and more  generally they can be associated to dynamical weights or fluxes associated to the links of the considered network. The considered coupling mechanism between topological signals of nodes and links is local, meaning that every node and every link is only affected by the dynamics of nearby nodes and links. In particular, the dynamics of the nodes is dictated by a Kuramoto-like system of equations where we introduce a phase lag that depends on the dynamical state of nearby links. Similarly, the dynamics of the links is dictated by a higher-order Kuramoto-like system of equations \cite{millan2020explosive} where we introduce a phase lag dependent on the dynamical state of nearby nodes.

On a fully connected network, Dirac synchronization is explosive as it leads to a discontinuous forward synchronization transition and a continuous backward synchronization transition.
Therefore, Dirac synchronization determines a topological mechanism to achieve abrupt discontinuous synchronization transitions. The theoretical investigation of the model predicts that the discontinuous transition occurs at a theoretically predicted value of the coupling constant $\sigma_c^{\star}$ when the incoherent phase loses stability. However, for smaller value of the coupling constant, the incoherent phase can coexist with the coherent one.

The coherent phase can be observed for $\sigma>\sigma_c$.
However, for $\sigma_c < \sigma<\sigma_S^{\star}$, the system is in a rhythmic phase characterized by non-stationary order parameters. Here, we theoretically predict the numerical value of $\sigma_S^{\star}$ and we investigate numerically the dynamics of the order parameters in the rhythmic phase. 

This work shows how topology can be combined with dynamical systems leading to a new framework to capture abrupt synchronization transitions and the emergence of non trivial rhythmic phases.

This work can be extended in different directions. First of all, the model can be applied to more complex network topologies including not only random graphs and scale-free networks but also real network topologies such as experimentally obtained brain networks. Secondly, using the higher-order Dirac operator \cite{bianconi2021topological,lloyd2016quantum}, Dirac synchronization can be extended to simplicial complexes where topological signals can be defined also on higher-order simplices such as triangles, tetrahedra and so on. We hope that this work will stimulate further theoretical and applied research along these lines.

\newpage
\section*{METHODS}
\subsection*{Internal frequencies of the projected dynamics}
\label{Ap0}
The frequencies $\hat{\bm\omega}$ characterising the uncoupled dynamics of the projected variables $\bm{\psi}$ can be determined using Eq. (\ref{hatomega}) from the internal frequencies of the links $\tilde{\bm\omega}$.
In particular, since the frequencies $\tilde{\bm\omega}$ are normally distributed, the frequencies $\hat{\bm\omega}$ will also be normally distributed. However, Eq.(\ref{hatomega}) implies that the frequencies $\hat{\bm\omega}$ are correlated.
Let us recall that the internal frequencies of the links $\tilde{\bm\omega}$ are taken to be  independent  Gaussian variables with zero average and standard deviation  $1/\sqrt{N-1}$, i.e. \bea\tilde{\omega}_{\ell}\sim \mathcal{N}(0,1/\sqrt{N-1})\eea for each link $\ell$ of the network.
Using the definition of the incidence matrix ${\bf B}$, it is easy to show that the expectation of $\hat{\omega}_i$ is given by  
\bea
\avg{\hat{\bm\omega}}={\bf B}\Avg{\bm{\tilde{\omega}}}={\bf 0}.
\label{avgomega}
\eea

Given that in a fully connected network each node has degree $k_i=N-1$, the covariance matrix ${\bf C}$ is given by the graph Laplacian ${\bf L}_{[0]}$ of the network, i.e.
 {\bea
C_{ij}&=&\Avg{\hat{\omega}_i\hat{\omega}_j}_c= \Avg{[{\bf B}\tilde{\bm\omega}]_i [{\bf B}\tilde{\bm\omega}]_j}_c\nonumber \\
&=&[{ L_{[0]}}]_{ij}\frac{1}{N-1},
\label{Cdef}
\eea}
where we have indicated with $\Avg{\ldots}_c$ the connected correlation.
Therefore, the covariance matrix has elements given by 
\bea
{\bf C}_{ij}=\delta_{ij}-\frac{1}{N-1}(1-\delta_{ij}).
\label{covariance}
\eea
Moreover   we  note that the average of $\hat{\bm\omega}$ over all the nodes of the network is zero. In fact
\bea
\sum_{i=1}^N \hat \omega_i = {\bf 1}^T \hat {\bm \omega} = {\bf 1}^T {\bf B}\tilde{\bm{\omega}} = 0,
\label{u}
\eea
where we indicate with ${\bf 1}$ the $N$-dimensional column vector of elements $1_i=1$.

With these hypotheses, the marginal probability  $G_1(\hat{\omega})$ that the internal frequency $\hat{\omega}_i$ of a generic node $i$ is given by $\hat{\omega}_i=\hat{\omega}$  can be expressed as (see \cite{ghorbanchian2020higher} for the derivation),
\bea
G_1(\hat{\omega})&=& \frac{1}{\sqrt{2\pi/\bar{c}}}\exp\left[-\bar{c}\frac{\hat{\omega}^2}{2}\right],
\label{marginal}
\eea
where we have put 
\bea
\bar{c}=\frac{N}{N-1}.
\eea

\subsection*{Derivation of the stationary state expression for $a_i$ and $b_i$}
\label{Ap1}
Let us now consider a given node $i$ and its continuity equation, Eq. (\ref{continuity}). By using the ansatz defined in Eq. (\ref{two}) and omitting the index $i$ as long as $X_{\alpha}\neq 0$, we can observe that the continuity Eq. (\ref{continuity}) is satisfied  if the following equations are satisfied 
for $m=0$, $n\neq 0$ and for $n=0$, $m\neq 0$, 
\begin{widetext}
 {\bea
na^{n-1}\partial_t a+\textrm{i}a^nn\kappa_{\alpha}+\frac{1}{2}\sigma n\left[X_{\alpha}a^{n+1}-X_{\alpha}^{\star}a^{n-1}\right]-\sigma \frac{1}{4}na^n(b-b^{\star})=0,\label{sysn0} \\
mb^{m-1}\partial_t b+\textrm{i}b^mm\kappa_{\beta}+\sigma \bar{c}\frac{1}{2}m\left[X_{\alpha}ab^m-X_{\alpha}^{\star}a^{\star}b^m\right]+\sigma \frac{1}{2}m(b^{m+1}-b^{m-1})=0,\label{sysm0}
\eea}
\end{widetext}
 Moreover, for every $m> 0$, $n> 0$ the following equation needs to be satisfied
\begin{widetext}
\bea
na^{n-1}b^m\partial_t a+ma^nb^{m-1}\partial_t b+\textrm{i}a^nb^mn\kappa_{\alpha}+\textrm{i}a^nb^mm\kappa_{\beta}\nonumber \\
+\frac{1}{2}\sigma n\left[X_{\alpha}a^{n+1}-X_{\alpha}^{\star}a^{n-1}\right]b^{m}
-\sigma \frac{1}{4}na^n(b^{m+1}-b^{m-1})\nonumber \\
+\sigma \bar{c}\frac{1}{2}m\left[X_{\alpha}a^{n+1}-X_{\alpha}^{\star}a^{n-1}\right]b^{m}
+\sigma \frac{1}{2}ma^n(b^{m+1}-b^{m-1})=0.
\label{sys}
\eea
\end{widetext}
As long as $a\neq 0$ and $X_{\alpha}\neq 0$, all these equations are satisfied   if and only if $|a|=|b|=1$ and 
\begin{widetext}
\bea
\partial_t a+\textrm{i}a\kappa_{\alpha}
+\frac{1}{2}\sigma \left[X_{\alpha}a^{2}-X_{\alpha}^{\star}\right]
-\sigma \frac{1}{4}a(b-b^{-1})=0,\nonumber \\
\partial_t b+\textrm{i}b\kappa_{\beta}
+\frac{1}{2}\sigma \bar{c}\left[X_{\alpha}a-X_{\alpha}^{\star}a^{-1}\right]b
+\sigma \frac{1}{2}(b^{2}-1)=0.
\label{dynA}
\eea
\end{widetext}
For $X_{\alpha}=0$, instead, by proceeding in a similar way, we get that for $|a|\neq 0$, $|b|=1$, the equations for $a$ and $b$ are given by 
\bea
&&\partial_t a+\textrm{i}a\kappa_{\alpha}
-\sigma \frac{1}{4}a(b-b^{-1})=0,\nonumber\\
&&\partial_t b+\textrm{i}b\kappa_{\beta}+\sigma \frac{1}{2}(b^{2}-1)=0.
\label{dynA0}
\eea
while for $|a|=0$, the equation for $b$ is unchanged but $b$ can have an arbitrary large absolute value.

\subsection*{The onset of the instability of the incoherent phase }
\label{crit}
In this Appendix we use stability considerations to derive the synchronization threshold $\sigma_{c}^{\star}$.  {This approach is analogous to similar approaches used to study the onset of instability of the incoherent phase in models focusing solely on signals defined on the nodes of a network \cite{peron2020collective,arola2022self}.}
Let us now consider the first of Eqs. (\ref{dyn2Ma}) for every $a_i(\omega_i,\hat{\omega}_i)$, and study the stability of the trivial solution in which $a_i(\omega_i,\hat{\omega}_i)=0$ for every node $i$ and every choice of the frequencies $(\omega,\hat{\omega})$, also implying that $R_{\alpha}=0$. The stationary solutions for $a_i$ and $b_i$ describe a discontinuous transition from steady state solutions with $a_i=0$ to $|a_i|=1$. Studying the stability of the incoherent phase is therefore non-trivial. In order to do so, we study the continuity equations using the generalized Ott-Antonsen ansatz and we consider non zero values of $a_i$ with absolute value $|a_i|\ll1$ while keeping $|b_i|=1$ so that $b_i^{-1}=b_i^{\star}$. With these hypotheses, we notice that Eq. (\ref{sys}) describing the dynamics of the high frequency $\alpha$ modes of $\rho(\alpha,\beta)$ are negligible and we can just focus on  Eq.~{\ref{sysn0}} obtained for $m=0,n=1$.
Since $a_i(\omega,\hat{\omega})$ is only a function of $(\omega,\hat{\omega})$, in this section we simplify the notation by omitting the index $i$. This entails for instance  considering  the function $a(\omega,\hat{\omega})$ instead of $a_i(\omega_i,\hat{\omega}_i)$. A similar convention is used for other variables only depending on the node $i$ through the internal frequencies $\omega=\omega_i$ and $\hat{\omega}=\hat{\omega}_i$. To this end, we write Eq. (\ref{sysn0}) as 
\bea
\partial_t a(\omega,\hat{\omega})=F(a(\omega,\hat{\omega}),b(\omega,\hat{\omega}),X_{\alpha}),
\label{1F}
\eea
with $F(a,b,X_{\alpha})$ given by 
\bea
F(a,b,X_{\alpha})&=&-\textrm{i}a\kappa_{\alpha}
-\frac{1}{2}\sigma \left[X_{\alpha}a^{2}-X_{\alpha}^{\star}\right]\nonumber \\
&&
+\sigma \frac{1}{4}a(b-b^{\star}),\nonumber
\eea
where $\kappa_{\alpha}=\kappa_{\alpha}(\omega,\hat{\omega})$.
In this equation $X_{\alpha}$ is intended to be a function of all the variables ${\bf a}$ according to Eqs. (\ref{defXbG}).
By linearizing Eq.~(\ref{1F})  for $a(\omega,\hat{\omega})=\Delta a(\omega,\hat{\omega})\ll 1$ for every value of $(\omega,\hat{\omega})$, and neglecting fluctuations in the variables $b(\omega,\hat{\omega})$, we obtain
\bea
\partial_t \Delta a(\omega,\hat{\omega})=[-\textrm{i}\kappa_{\alpha}+\sigma B]\Delta a+\sigma \mathcal{S}/2,
\label{1Fline}
\eea
where we have defined $B$ as
\bea
B=\frac{1}{4} (\bar{b}-\bar{b}^{\star}),
\eea
with $\bar{b}=\bar{b}(\omega,\hat{\omega})$ being the stationary solution of Eq. (\ref{dyn2Ma}) in the limit $X_{\alpha}\to 0$ and $a_j\to 0,\ \forall j$  and where  $\mathcal{S}$  indicates 
\bea
\mathcal{S}=\int d\omega' \int d\hat{\omega}'G_0(\omega')G_1(\hat{\omega'})\Delta a(\omega',\hat{\omega}').
\label{Sdef}
\eea
In order to predict the onset of the instability of the incoherent phase, i.e., in order to predict the value of $\sigma_c^{\star}$, we study the discrete spectrum of Eq.~(\ref{1Fline}). Assuming that Eq.~(\ref{1Fline}) has Lyapunov exponent $\lambda$, we find that  $\Delta a(\omega,\hat{\omega})$ obeys
\bea
\Delta a(\omega,\hat{\omega})=\frac{1}{2}\mathcal{S} \hat\Delta(\omega,\hat{\omega}).
\label{ai2}
\eea
where $ {\hat\Delta}(\omega,\hat{\omega})$ is given by 
\bea
 {\hat\Delta}(\omega,\hat{\omega})&=&\frac{1}{(\textrm{i}\kappa_{\alpha}+\lambda)/\sigma- B},
 \label{Delta}
\eea

By inserting Eq.~(\ref{ai2}) in the definition of $\mathcal{S}$ given by Eq.~(\ref{Sdef}),  we obtain a self-consistent equation that reads
\bea
1=\frac{{1}}{2}\hat{I}=\frac{1}{2}\int d\omega' \int d\hat{\omega}'G_0(\omega')G_1(\hat{\omega'})\hat{\Delta}(\omega',\hat{\omega}').
\label{self1}
\eea

Therefore, this equation provides the value of the Lyapunov exponent $\lambda$ for any given value of the coupling constant $\sigma$. 
We look for the onset of the instability $\sigma=\sigma_c^{\star}$ of 
the incoherent  solution $R_{\alpha}=0$  by imposing that its Lyapunov exponent vanishes, i.e., $\lambda=0$.

In order to solve this equation, we need to find the explicit form for $B$ in the limit $X_{\alpha}\to 0$. 
By considering the stationary solution in the incoherent phase, we obtain that the variable $B$ is given by 
\bea
B(\omega,\hat{\omega})=-\textrm{i}\frac{\kappa_{\beta}}{2\sigma}.
\eea
as long as $|b_i|=1$, i.e. as long as $|\kappa_{\beta}/\sigma|\leq 1$.  We can now insert this expression in $\hat\Delta$ finding
\bea
\hat\Delta^{-1}=\textrm{i}\frac{\kappa_{\alpha}+\kappa_{\beta}/2}{\sigma}=\textrm{i}\frac{\omega-\hat{\Omega}}{\sigma}\left({1+\frac{1}{2}\bar{c}}\right),
\eea
for $|\kappa_{\beta}/\sigma|\leq 1$.
Otherwise, since we assume $a_i\neq 0$, $b_i$ does not have a stationary value and in expectation over time we have 
\bea
\overline{B(\omega,\hat{\omega})}=0.
\eea
This leads to 
\bea
\hat\Delta^{-1}=\textrm{i}\frac{\kappa_{\alpha}}{\sigma},
\eea
for $|\kappa_{\beta}/\sigma|\geq 1$.

In the limit $N\to \infty$, we have that $\bar{c}\to 1$ and $\hat{\Omega}\to\Omega_0.$ Therefore in this limit, we obtain 
\bea
\hspace*{-8mm}\hat\Delta^{-1}=\left\{\begin{array}{ccc}\textrm{i}({3}/{2}){(\omega-{\Omega}_0)}/{\sigma}&\mbox{for}&|\kappa_{\beta}/\sigma|\leq 1,\\ \textrm{i}{(\omega-{\Omega}_0+\hat{\omega}/2)}/{\sigma}&\mbox{for} &|\kappa_{\beta}/\sigma|\geq 1.\end{array}\right.
\eea
By using this explicit expression for $\hat\Delta$ in terms of the frequency of the nodes, it is straightforward  to see that  $\hat{I}$   can be evaluated with the method of residues leading to 
\bea
\hat{I}=\sqrt{\frac{\pi}{2}}\sigma\left[\frac{2}{3} \mbox{erf}\left(\frac{\sigma}{2}\right)+\frac{2}{\sqrt{5}}\mbox{erfc}\left(\frac{1}{3}\sqrt{\frac{5}{2}\sigma}\right)\right].
\eea
By inserting the values of these integrals in Eq. (\ref{self1}) and solving for $\sigma$,  we get that the synchronization threshold occurs at
\bea
\sigma_c^{\star}=2.14623\ldots.
\eea

\section*{Acknowledgements}
G. B. acknowledges funding from the Alan Turing Institute and from Royal Society  IEC\textbackslash NSFC\textbackslash191147.  J.J.T. acknowledges financial support from the Consejería de Transformación Económica, Industria, Conocimiento y Universidades, Junta de Andalucía and European Regional Development Funds, Ref. P20\_00173. This work is also part of the Project of I+D+i Ref. PID2020-113681GB-I00, financed by MICIN/AEI/10.13039/501100011033 and FEDER “A way to make Europe”.
This research utilized Queen Mary's Apocrita HPC facility, supported by QMUL Research-IT. http://doi.org/10.5281/zenodo.438045. 

\bibliography{references}
\end{document}